\documentclass[aps,superscriptaddress,eqsecnum,nofootinbib,preprintnumbers]{revtex4}
\usepackage{graphicx,epsfig}
\usepackage{amsmath}
\usepackage{amssymb}
\usepackage{subfigure}

\usepackage{xfrac}

\usepackage{accents}

\usepackage{relsize}

\usepackage[normalem]{ulem}

\usepackage[utf8]{inputenc}
\usepackage[english]{babel}
\usepackage[dvipsnames]{xcolor}

\newcommand{\nn}{\nonumber}

\newcommand{\dispdot}[2][.2ex]{\dot{\raisebox{0pt}[\dimexpr\height+#1][\depth]{$#2$}}}

\newcommand\myprime{\mkern-0.1mu\raise1.4ex\hbox{$\scriptstyle\prime$}}

\newcommand*{\LargerCdot}{\raisebox{-0.25ex}{\scalebox{1.1}{$\cdot$}}}

\begin{document}

\title{Effective field equations and scale-dependent couplings in gravity}

\date{\today}

\author{Alfio Bonanno}
\email{alfio.bonanno@inaf.it}
\affiliation{INAF, Osservatorio Astrofisico di Catania, Via S.Sofia 78, 95123 Catania, Italy}
\affiliation{INFN, Sezione di Catania, Via S. Sofia 72, 95123 Catania, Italy}

\author{Georgios Kofinas}
\email{gkofinas@aegean.gr}
\affiliation{Research Group of Geometry, Dynamical Systems and Cosmology,\\
Department of Information and Communication Systems Engineering,\\
University of the Aegean, Karlovassi 83200, Samos, Greece}

\author{Vasilios Zarikas}
\email{vzarikas@uth.gr}
\affiliation{Nazarbayev University, School of Engineering, Astana, Republic of Kazakhstan, 010000}

\begin{abstract}
A new set of field equations for a space-time dependent Newton's constant $G(x)$ and cosmological
constant $\Lambda(x)$ in the presence of matter is presented. We prove that it represents the
most general mathematically consistent, physically plausible, set of evolution equations assuming at
most second derivatives in the dynamical variables. In the new Einstein's equations, only
$\Lambda$-kinetic terms arise, while in the modified conservation equation, derivative terms of $G$
also appear. As an application, this formalism is applied in the context of the Asymptotic Safety
scenario to the early universe, assuming a perfect fluid with a radiation equation of state.
Cosmological solutions are obtained for all types of spatial curvature, displaying a variety of
interesting cosmic evolutions. As an indication of such behaviours, bouncing solutions, recollapsing
solutions or non-singular expanding solutions with a transient acceleration era are discussed in
details.
\end{abstract}

\maketitle

\section{Introduction}
\label{Introduction}

The old idea \cite{milne37} that the gravitational constant $G$ is a space-time dependent quantity
has recently received much attention. In the original formulation by Dirac \cite{dirac37}, the secular
variation of Newton constant was postulated within the Large Number Hypothesis \cite{dirac38}, but
it has since then  gained support over the years in various contexts (see \cite{kragh2016varying} for
a recent review on the subject). On the other hand, since the early development of the theory, the
issue of the mathematical compatibility with Einstein theory of gravity was clearly recognized as
a crucial one, because in this case diffeomorphism invariance is immediately broken. In particular
one is forced to introduce additional compatibility laws in order to preserve Bianchi identities
and relations of the type $(G(x^\sigma) T^{\mu\nu})_{; \nu}=0$ must artificially be assumed,
being $T^{\mu\nu}$  the energy-momentum tensor of the matter component (here the semicolon denotes
the covariant derivative). A natural way out to this impasse is to promote $G$ to the status of a
dynamical variable, as in the seminal work by Brans and Dicke \cite{brans61}. In this case,
$G\propto 1/\Phi$  where $\Phi$ is a scalar field whose Lagrangian resembles a free scalar field
with a non-trivial derivative coupling. Albeit scalar-tensor theories are considered viable
alternative to Einstein theory of gravitation, a prescribed physical mechanism for variable
gravitational constant is usually not allowed, as the evolution of $\Phi$ is dictated by its own
equation of motion whose precise form is rather arbitrary: one is free to add any possible
self-interaction term to the $\Phi$-field  Lagrangian.

In recent times the idea that the coupling constants of certain field theories might become
functions of spacetime coordinates has emerged both in classical and in quantum mechanical
contexts (see \cite{uzan11} for a review of various models and various constraints).

In \cite{dvali11} it has been shown that the {\it classical} $\lambda \phi^4$ theory, in presence
of an external source, develops a non-trivial scaling property as a function of an UV cutoff which
is represented by the (inverse) of the characteristic distance $\ell$  at which the screening
effect of the charge is present. This model exhibits dimensional transmutation and asymptotic
freedom depending on the sign of $\lambda$. The important lesson to be learned from this
calculation is that the non-trivial scaling already occurs at the level of the equation of motion
and has an entirely classical origin.

Another type of classical scaling of Newton constant and the cosmological constant might also
emerge via the so called covariant averaging procedure \cite{mauro95}. In this case both $G$ and
$\Lambda$ must acquire a scale dependence if one considers a density distribution which is
inhomogeneous at small distances and then performs spatial averaging over 3-volumes of increasing
linear extension $\ell$. As in the previous example, the classical dynamics of the  averaged
quantities is encoded in a spacetime dependence of $G$ and $\Lambda$, which is ruled by a
{\it classical} renormalization group (see also \cite{brande}, \cite{unruh98}).

Various approaches to  Quantum Gravity (QG) lead to a scale-dependent Newton constant albeit with
different scaling law. In the Effective Field Theory approach \cite{1994PhRvD..50.3874D}, the actual
scaling is obtained by Fourier transforming (thus assuming $k\propto 1/r$ where $k$ is the modulus
of the 3-momentum) the low-energy limit of the scattering amplitude of two heavy particles, so that
their renormalized Newtonian potential becomes a distance-dependent Newton constant of the
type $G_{\rm eff}(r)=G (1+41 G/ 10 \pi r^2 )$ \cite{bborh2003}. Conversely, in the lattice inspired
approach of  \cite{hamber07}, the momentum dependence of Newton constant leads to  functional
forms of the type $G=G(\Box)$, where $\Box$ is the covariant Laplacian. In all these cases one
expects to deal with a new class of effective field equations encoding the ``running'' of $G$  in
a robust mathematical formalism.

In the Asymptotic Safety (AS) scenario for QG \cite{1979grec.conf..790W}-\cite{1999PThPh.102..181S},
the search for a consistent continuum limit of gravity is obtained by means of a covariant
formulation of the Wilsonian renormalization group (RG) \cite{Berges:2000ew}-\cite{Gies:2006wv}.
An effective action can be calculated using RG functional renormalisation and heat kernel
techniques \cite{Codello:2015oqa}, \cite{Zarikas:1999bf}. We refer the reader to textbooks
\cite{percacci17}, \cite{reuter2019quantum} and the reviews
\cite{2011RSPTA.369.2759L}-\cite{2020FrP.....8..269B} for a more detailed discussion.

In more detail, at variance with the well-known renormalization procedure in quantum-field theory,
in this approach the removal of the cutoff $k$ is obtained in a statistical-mechanical sense,
i.e. by assuming a second-order phase transition in the ultraviolet critical surface. The role
of the cutoff is therefore deeply connected with the block-spin transformation, or Kadanoff
blocking, a well-known procedure in lattice field theory and condensed matter \cite{Wilson:1973jj}.
According to this idea, the cutoff $k$ has not the meaning of a momentum exchanged in a scattering
process, neither the RG scale often used in dimensional regularization in effective field theory
approaches to QG \cite{don}. On the contrary, if one assumes Fourier transformability of the
block-spin transformation, $k$ turns out to be  roughly of the order of the inverse of the lattice
size.  One can imagine that as $k\rightarrow 0$ the size of the characteristic box where the physics
is probed  tends to infinity. As a consequence, various scaling behaviors for $G(k)$ have been
proposed in astrophysical \cite{Bonanno:2000ep}, \cite{Bonanno:2019ilz} and cosmological contexts
\cite{Bonanno:2007wg}, \cite{2016PhRvD..94j3514K} (see \cite{Bonanno:2017pkg} for a review).

In spite of the above efforts, no well established  formalism to embed a generic running
of the type $G=G(x^\mu)$, $\Lambda=\Lambda(x^{\mu})$ in a modification of standard Einstein's
relativity is available. A spacetime dependence of the gravitational constant has in fact the
status of an external field and its incorporation in a Lagrangian \cite{Reuter:2003ca} or
Hamiltonian \cite{Bonanno:2004ki}, \cite{Bonanno:2017gji} formalism is problematic (see also
\cite{smolin2015} for an alternative approach, and \cite{2020ApJ...893L..35B} for a recent discussion
of possible variation of $G$ on secular time-scales).

The purpose of the approach in the present work is to describe a possible formulation based on a
new set of modified effective equations of motion. The advantage of this approach will soon
become apparent because, at variance with the framework developed in \cite{Reuter:2003ca}, no
non-standard contributions to the stress-energy tensor appear in this case. We here present in
detail the derivation of the new set of effective equations of motion and discuss some application
in cosmology.

The plan of the paper is the following: In section II, the new field equations with
spacetime-dependent cosmological constant $\Lambda(x^{\mu})$ and Newton constant $G(x^{\mu})$ in
the presence of matter are derived, and a thorough discussion and analysis of the possible options
is given. Section III presents the application of these field equations to the case of a
cosmological metric with a fluid as the matter component, without further specifying the form of
the fluid or identifying the underlying theory of varying constants. In Section IV, the Asymptotic
Safety scenario is used as an instance for the identification of the form of $\Lambda,G$ in
the early universe and the full cosmological evolution is obtained for a radiation equation
of state. Finally, Section V presents the overall conclusions.

\section{Field equations with varying $G$ and $\Lambda$ in the presence of matter}
\label{nkjk}

Let us consider a  spacetime-dependent cosmological constant $\Lambda(x)$ and Newton constant
$G(x)$ in the 4-dimensional Einstein equations with a possible matter energy-momentum tensor
$T_{\mu\nu}$, i.e. $G_{\mu\nu}=-\Lambda(x) g_{\mu\nu}+8\pi G(x) T_{\mu\nu}$. Bianchi identities
$G_{\mu\nu}{^{;\mu}}=0$ imply the consistency condition \cite{Bonanno:2001hi}
\begin{equation}
8\pi(GT_{\mu\nu}){^{;\mu}}=\Lambda_{;\nu}\,,
\label{alf}
\end{equation}
which can only be satisfied in general either assuming an exchange of energy between the matter
component and the variation of $\Lambda$ \cite{Bonanno:2007wg}, or with the more stringent
condition $8\pi T_{\mu\nu}G^{;\mu}=\Lambda_{;\nu}$ if the energy momentum tensor is strictly
conserved. In particular in vacuum, equation (\ref{alf}) forces to assume a constant $\Lambda$, as
it would therefore be impossible to consider a variable $\Lambda$ theory \cite{Kofinas:2015sna} in
this case.

In order to overcome the impasse of previous works, one can instead add to the Einstein equations
appropriate terms which depend on the objects at hand (the functions $\Lambda, G$ and $T_{\mu\nu}$
(and certainly $g_{\mu\nu}$)) and ask for mathematical and physical consistency. We are not satisfied
to only find consistent equations, but we also intend to investigate in which directions the derived
equations decline to be the most general construction. Our tool in this quest will be the exhaustive
treatment of the consistency conditions arising from the Bianchi identities. The quantities
$\Lambda, G$ have been promoted from constants to spacetime functions. In the special case where
$\Lambda, G$ become constants, the Einstein equations should remain intact (actually, not even a
simple rescaling of these constant values $\Lambda, G$ is allowed). Therefore, the appropriate terms
to be added must have the property that when $\Lambda, G$ become constants, these extra terms vanish.
This condition is only satisfied if each of the added terms contains at least one derivative of
$\Lambda$ or $G$. Thus, the added terms should be, what can be called, kinetic terms of $\Lambda$ or
$G$. To state it slightly differently, spacetime dependent rescalings of $G_{\mu\nu}$, $g_{\mu\nu}$,
$T_{\mu\nu}$ could be added once, and then, all the accumulated effects of these rescalings form
our $\Lambda, G$ where we start from. To become more concrete, we find it convenient to
parametrize $\Lambda,G$ by introducing the dimensionless quantities $\psi(x),\chi(x)$ as
\begin{equation}
\Lambda=\bar{\Lambda}\,e^{\psi}\,\,\,\,\,\,,\,\,\,\,\,\,G=\bar{G}\,e^{\chi}\,,
\label{asjn}
\end{equation}
where $\bar{\Lambda}$ (with dimensions inverse length squared) and $\bar{G}$ (with dimensions
length squared) are arbitrary constant reference values. Note that (\ref{asjn}) does not allow
for a change of sign of $G$ or $\Lambda$. While it is reasonable to suppose that $G$ is always
positive, this is not generally the case for $\Lambda$. Albeit our formalism is very general, we
shall be mostly interested considering scaling laws for $G$ and $\Lambda$ which are valid near a
fixed-point, so that the scaling behavior is already defined and no change of sign is possible.

The kinetic terms should contain at least one derivative of $\psi$ or $\chi$. As a result, terms
such as $\Lambda A(\psi)g_{\mu\nu}$, $A(\psi)R_{\mu\nu}$, $\widetilde{A}(\chi)Rg_{\mu\nu}$,
$G\widetilde{A}(\chi)R_{\kappa(\mu}R^{\kappa}_{\,\,\,\nu)}$, $GA(\psi)R^{2}g_{\mu\nu}$, ...
(or terms that arise from an action and ensure the Bianchi identities on their own)
cannot be added because these terms are not deviations of Einstein gravity and for $\psi,\chi$
constants, they do not reduce to Einstein gravity but to another gravity theory.
Similarly, terms containing $T_{\mu\nu}$, such as
$G\widetilde{A}(\chi)T_{\mu\nu}$, $GA(\psi)Tg_{\mu\nu}$,
$G^{3}\widetilde{A}(\chi)T_{\kappa(\mu}T^{\kappa}_{\,\,\,\nu)}$, $G^{3}A(\psi)T^{2}g_{\mu\nu}$,
$G^{3}\widetilde{A}(\chi)TT_{\mu\nu}$, $G^{2}A(\psi)R_{\kappa(\mu}T^{\kappa}_{\,\,\,\nu)}$, ...
are not added since they form undesirable deviation from Einstein gravity. Note that the case
where $G$ is a derivative operator $G(\Box)$, i.e. a function of d'Alembertian, starting from a
constant value and corrected by powers of $\Box$ or non-local \cite{Hamber:2006rq} powers of $\Box$,
is also permitted. Such a term is still of the form $GT_{\mu\nu}$, namely
$\bar{G}[1+A(\Box)]T_{\mu\nu}$, and the only difference is that $G$ is not a spacetime function, but
creates derivatives of $T_{\mu\nu}$, therefore kinetic terms of $G$ cannot be added now.
This case will also be considered aside in the following analysis.

One difference between the functions $\Lambda,G$ on one side and $T_{\mu\nu}$ on the other side is
that $\Lambda,G$ can be considered as known/predefined spacetime functions (which do not obey some
equations of motion), while $T_{\mu\nu}$ is governed by some equations of motion (conservation
equations) which will actually arise as usually from the Bianchi identities
themselves. Note that there is still the reasonable option that $\Lambda,G$ are related to some
proper distance in spacetime, which is therefore given by an integral, and then differential
equations arise for $\Lambda,G$; however, these differential equations are not under quest, and
moreover, since we want to have a uniform treatment of this case along with the simpler situation
where $\Lambda,G$ are manifest spacetime functions, we will consider $\Lambda,G$ in the elaboration
of the Bianchi identities as given spacetime functions. In Asymptotic Safety in particular (as
opposed to other varying constants theories of gravity and cosmology), the quantities $\Lambda,G$ are
functions of the energy scale $k$, and if $k$ is attached to a spacetime function ($k$ transforms as
a scalar under coordinate changes), then $\Lambda, G$ become spacetime functions;
usually, $k$ depends on a single spacetime coordinate and so $\Lambda,G$
become functions of this coordinate. We will not stick in this analysis to the special case of AS,
but we will consider $\Lambda,G$ arbitrary spacetime functions. Due to the above distinct feature
of $\Lambda, G$ and $T_{\mu\nu}$, the terms to be added in Einstein equations will be tensorial
quantities constructed solely out of $\Lambda,G$ and their derivatives, so no new added term or
coefficient should contain $T_{\mu\nu}$. Thus, kinetic terms including $T_{\mu\nu}$ multiplied by
$\psi_{;\mu}$ or $\chi_{;\mu}$, such as $G^{2}A(\psi)T_{\kappa(\mu}\psi_{;\nu)}\psi^{;\kappa}$,
$G^{2}\widetilde{A}(\chi)Tg_{\mu\nu}\psi^{;\kappa}\psi_{;\kappa}$,
$G^{2}A(\psi)T\chi_{;\mu}\chi_{;\nu}$, ..., although they reduce to Einstein gravity for $\psi, \chi$
constants, however from the Bianchi identities it is not possible to determine the various coefficients
$A,\widetilde{A}$ since any such term will contribute to the conservation equation for matter and
not to some other cancelation needed for consistency (we will come back again on this issue of
cancelation in the following and it will become more clear). This means that there is no preference
on any particular such term containing $T_{\mu\nu}$, among the infinity of similar terms, and
therefore none of these terms should be added in the Einstein equations if one is to construct a
meaningful and well-motivated physical theory.

Out of the previous discussion, what only remains for constructing the modified Einstein equations
are kinetic terms such as $\psi_{;\mu}\psi_{;\nu}$, $g_{\mu\nu}\Box\psi$,
$g_{\mu\nu}\psi^{;\rho}\chi_{;\rho}$, $\chi_{;\mu;\nu}$, ...,
$G\psi^{;\kappa}\psi_{;\kappa}\psi_{;\mu}\psi_{;\nu}$, $G\psi_{;\mu}\psi_{;\nu}\Box{\chi}$,
$G\chi_{;\kappa;(\mu}\chi_{;\nu)}\chi^{;\kappa}$, ..., each one with a dimensionless coefficient
being in general function of both $\psi,\chi$. These terms contain an even number of derivatives
(two, four or higher) of $\psi$ or $\chi$. Other terms with four or more derivatives, containing
the Ricci or the Riemann tensor, such as $GR_{\kappa(\mu}\psi_{;\nu)}\psi^{;\kappa}$,
$GRg_{\mu\nu}\chi^{;\kappa}\chi_{;\kappa}$, ..., each one with a dimensionless coefficient, are
not permissible since they do not provide meaningful equations of motion (this will be realized
later through the derivation procedure). Our assumption in this work is the restriction to the use
of up to second derivatives in the equations of motion, so as $G_{\mu\nu}$ contains second
derivatives of the metric tensor, similarly the extra kinetic terms to be added in the Einstein
equations will have second derivatives, which will thus be derivatives of $\psi,\chi$.
One could attempt in the future to include kinetic terms with four derivatives as well. Therefore,
in the present work we will consider the terms $\psi_{;\mu}\psi_{;\nu}$,
$g_{\mu\nu}\psi^{;\rho}\psi_{;\rho}$, $\psi_{;\mu;\nu}$, $g_{\mu\nu}\Box\psi$, each one with a
different coefficient $\mathcal{A}(\psi,\chi)$, ..., also the terms $\chi_{;\mu}\chi_{;\nu}$,
$g_{\mu\nu}\chi^{;\rho}\chi_{;\rho}$, $\chi_{;\mu;\nu}$, $g_{\mu\nu}\Box\chi$, again each one with
its own coefficient $\widetilde{\mathcal{A}}(\psi,\chi)$, ..., and finally the mixed terms
$\psi_{;(\mu}\chi_{;\nu)}$, $g_{\mu\nu}\psi^{;\rho}\chi_{;\rho}$ with their own coefficients too.
Bianchi identities will provide in principle partial differential equations for the various
coefficients, which fortunately however, as will be seen in the following analysis, will be reduced
to simple algebraic equations, from where all these coefficients will turn out to get unique numerical
values.

To this end, we add to the Einstein equations the energy-momentum tensor $\vartheta_{\mu\nu}(\psi)$
containing all possible kinetic terms of $\psi$, the energy-momentum tensor
$\widetilde{\vartheta}_{\mu\nu}(\chi)$ containing all possible kinetic terms of $\chi$ and the
only possible mixed terms $\psi_{;(\mu}\chi_{;\nu)}$, $g_{\mu\nu}\psi^{;\rho}\chi_{;\rho}$. Thus,
\begin{equation}
G_{\mu\nu}=-\Lambda g_{\mu\nu}+\vartheta_{\mu\nu}+\widetilde{\vartheta}_{\mu\nu}
+\mathcal{F}(\psi_{;\mu}\chi_{;\nu}+\psi_{;\nu}\chi_{;\mu})
+\mathcal{H}g_{\mu\nu}\psi^{;\rho}\chi_{;\rho}+8\pi G T_{\mu\nu}\,,
\label{njgc}
\end{equation}
where
\begin{equation}
\vartheta_{\mu\nu}=\mathcal{A}\,\psi_{;\mu}\psi_{;\nu}+\mathcal{B}\,g_{\mu\nu}\psi^{;\rho}
\psi_{;\rho}+\mathcal{C}\,\psi_{;\mu;\nu}+\mathcal{E}\,g_{\mu\nu}\Box\psi\,,
\label{fsdmn}
\end{equation}
\begin{equation}
\widetilde{\vartheta}_{\mu\nu}=\widetilde{\mathcal{A}}\,\chi_{;\mu}\chi_{;\nu}+
\widetilde{\mathcal{B}}\,g_{\mu\nu}\chi^{;\rho}\chi_{;\rho}
+\widetilde{\mathcal{C}}\,\chi_{;\mu;\nu}+\widetilde{\mathcal{E}}\,g_{\mu\nu}\Box\chi
\label{fsdjn}
\end{equation}
and all coefficients $\mathcal{A}, \mathcal{B}, \mathcal{C}, \mathcal{E}, \widetilde{\mathcal{A}},
\widetilde{\mathcal{B}}, \widetilde{\mathcal{C}}, \widetilde{\mathcal{E}}, \mathcal{F},
\mathcal{H}$ are functions of both $\psi, \chi$.
Here, $\Box\psi=\psi_{;\mu}^{\,\,\,\,\,;\mu}$, $\Box\chi=\chi_{;\mu}^{\,\,\,\,\,;\mu}$ and
$;$ denotes covariant differentiation with respect to the Levi-Civita connection of $g_{\mu\nu}$.
The traces $\vartheta=\vartheta^{\mu}_{\,\,\,\mu}$\,,\,$\widetilde{\vartheta}=
\widetilde{\vartheta}^{\mu}_{\,\,\,\mu}$ take the forms
\begin{equation}
\vartheta=(\mathcal{A}+4\mathcal{B})\psi^{;\mu}\psi_{;\mu}+(\mathcal{C}+4\mathcal{E})\Box\psi\,,
\label{jjs}
\end{equation}
\begin{equation}
\widetilde{\vartheta}=(\widetilde{\mathcal{A}}+4\widetilde{\mathcal{B}})\chi^{;\mu}\chi_{;\mu}
+(\widetilde{\mathcal{C}}+4\widetilde{\mathcal{E}})\Box\chi\,.
\label{jgrs}
\end{equation}
Equation (\ref{njgc}) is also written as
\begin{eqnarray}
\!\!\!\!\!R_{\mu\nu}&\!=\!&\Lambda g_{\mu\nu}+\vartheta_{\mu\nu}-\frac{1}{2}\vartheta\,g_{\mu\nu}
+\widetilde{\vartheta}_{\mu\nu}-\frac{1}{2}\widetilde{\vartheta}\,g_{\mu\nu}
+\mathcal{F}(\psi_{;\mu}\chi_{;\nu}+\psi_{;\nu}\chi_{;\mu})
-(\mathcal{F}+\mathcal{H})g_{\mu\nu}\psi^{;\rho}\chi_{;\rho}
+8\pi G\Big(\!T_{\mu\nu}\!-\!\frac{1}{2}Tg_{\mu\nu}\!\Big)\label{ajdv}\\
&\!=\!&\mathcal{A}\,\psi_{;\mu}\psi_{;\nu}-\Big(\frac{\mathcal{A}}{2}+\mathcal{B}\Big)
g_{\mu\nu}\psi^{;\rho}\psi_{;\rho}
+\mathcal{C}\psi_{;\mu;\nu}-\Big(\frac{\mathcal{C}}{2}+\mathcal{E}\Big)g_{\mu\nu}\Box\psi+
\bar{\Lambda}e^{\psi}g_{\mu\nu}\nn\\
&&+\,\widetilde{\mathcal{A}}\,\chi_{;\mu}\chi_{;\nu}-\Big(\frac{\widetilde{\mathcal{A}}}{2}
+\widetilde{\mathcal{B}}\Big)
g_{\mu\nu}\chi^{;\rho}\chi_{;\rho}+\widetilde{\mathcal{C}}\chi_{;\mu;\nu}-
\Big(\frac{\widetilde{\mathcal{C}}}{2}+\widetilde{\mathcal{E}}\Big)g_{\mu\nu}\Box\chi\nn\\
&&+\mathcal{F}(\psi_{;\mu}\chi_{;\nu}+\psi_{;\nu}\chi_{;\mu})
-(\mathcal{F}+\mathcal{H})g_{\mu\nu}\psi^{;\rho}\chi_{;\rho}
+8\pi G\Big(\!T_{\mu\nu}\!-\!\frac{1}{2}Tg_{\mu\nu}\!\Big)\,,
\label{hshk}
\end{eqnarray}
while the Ricci scalar gets the form
\begin{eqnarray}
R&\!=\!&4\Lambda-\vartheta-\widetilde{\vartheta}-(2\mathcal{F}+4\mathcal{H})
\psi^{;\mu}\chi_{;\mu}-8\pi GT\label{jwuf}\\
&\!=\!&4\bar{\Lambda}e^{\psi}-(\mathcal{A}\!+\!4\mathcal{B})\psi^{;\mu}\psi_{;\mu}-
(\mathcal{C}\!+\!4\mathcal{E})\Box{\psi}
-(\widetilde{\mathcal{A}}\!+\!4\widetilde{\mathcal{B}})\chi^{;\mu}\chi_{;\mu}-
(\widetilde{\mathcal{C}}\!+\!4\widetilde{\mathcal{E}})\Box{\chi}
-(2\mathcal{F}\!+\!4\mathcal{H})\psi^{;\mu}\chi_{;\mu}-8\pi GT\,,\label{mjue}
\end{eqnarray}
where $T=T^{\mu}_{\,\,\,\mu}$.

The Bianchi identities that arise from (\ref{njgc}) are
\begin{eqnarray}
&&\vartheta_{\mu\nu}^{\,\,\,\,\,\,\,;\mu}+\widetilde{\vartheta}_{\mu\nu}^{\,\,\,\,\,\,\,;\mu}
-\bar{\Lambda}e^{\psi}\psi_{;\nu}
+\mathcal{F}\hspace{0.2mm}'\psi^{;\mu}\psi_{;\mu}\chi_{;\nu}
+\,\,\dispdot[.1ex]{\!\!\mathcal{F}}\chi^{;\mu}\chi_{;\mu}\psi_{;\nu}
+(\mathcal{F}\hspace{0.2mm}'\!+\mathcal{H}\hspace{0.15mm}')\psi^{;\mu}\chi_{;\mu}\psi_{;\nu}
+(\,\,\dispdot[.1ex]{\!\!\mathcal{F}}+\,\dispdot[.1ex]{\!\!\mathcal{H}})
\psi^{;\mu}\chi_{;\mu}\chi_{;\nu}\nn\\
&&+\mathcal{F}\,\Box{\psi}\,\chi_{;\nu}+\mathcal{F}\,\Box{\chi}\,\psi_{;\nu}
+(\mathcal{F}+\mathcal{H})\psi^{;\mu}\chi_{;\mu;\nu}
+(\mathcal{F}+\mathcal{H})\chi^{;\mu}\psi_{;\mu;\nu}
+8\pi\bar{G}e^{\chi}(T_{\mu\nu}{^{;\mu}}+T_{\mu\nu}\chi^{;\mu})
=0\,,\label{swsbn}
\end{eqnarray}
where a prime denotes partial differentiation with respect to $\psi$ (e.g.
$\mathcal{F}\hspace{0.2mm}'=\partial\mathcal{F}/\partial\psi$)
and a dot partial differentiation with respect to $\chi$ (e.g.
$\dispdot[.1ex]{\!\!\mathcal{F}}=\partial\mathcal{F}/\partial\chi$).
In the case of $G(\Box)$, the previous equations up to (\ref{mjue}) remain the same except
that the terms with derivatives of $\chi$ disappear; in equation (\ref{swsbn}) the last
two terms containing matter are substituted by $8\pi(GT_{\mu\nu})^{;\mu}$.

We can straightforwardly compute from (\ref{fsdmn}), (\ref{fsdjn}) the covariant derivatives
present in (\ref{swsbn}) as
\begin{eqnarray}
\vartheta_{\mu\nu}^{\,\,\,\,\,\,\,;\mu}&=&
(\mathcal{A}'\!+\!\mathcal{B}\hspace{0.1mm}')\psi^{;\mu}\psi_{;\mu}\psi_{;\nu}
+(\mathcal{E}\hspace{0.1mm}'\!+\!\mathcal{A})\Box\psi\,\psi_{;\nu}
+(\mathcal{C}'\!+\!\mathcal{A}\!+\!2\mathcal{B})
\psi^{;\mu}\psi_{;\mu;\nu}+\mathcal{C}\,\Box(\psi_{;\nu})+\mathcal{E}(\Box\psi)_{;\nu}\nn\\
&&+\,\,\,\dispdot[.1ex]{\!\!\mathcal{A}}\,\psi^{;\mu}\chi_{;\mu}\psi_{;\nu}
+\,\,\dispdot[.1ex]{\!\!\mathcal{B}}\,\psi^{;\mu}\psi_{;\mu}\chi_{;\nu}
+\,\,\dispdot[.1ex]{\!\!\mathcal{E}}\,\Box\psi\,\chi_{;\nu}
+\,\,\dispdot[.1ex]{\!\!\mathcal{C}}\,\chi^{;\mu}\psi_{;\mu;\nu}\,,
\label{wksh}
\end{eqnarray}
\begin{eqnarray}
\widetilde{\vartheta}_{\mu\nu}^{\,\,\,\,\,\,\,;\mu}&=&
(\,\,\dispdot[.1ex]{\!\!\widetilde{\mathcal{A}}}\!+\,
\dot{\!\!\widetilde{\mathcal{B}}}\,)\chi^{;\mu}\chi_{;\mu}\chi_{;\nu}
+(\,\,\dot{\!\!\widetilde{\mathcal{E}}}\!+\!\widetilde{\mathcal{A}}\,)\Box\chi\,\chi_{;\nu}
+(\,\,\dot{\!\!\widetilde{\mathcal{C}}}\!+\!
\widetilde{\mathcal{A}}\!+\!2\widetilde{\mathcal{B}}\,)\chi^{;\mu}\chi_{;\mu;\nu}
+\widetilde{\mathcal{C}}\,\Box(\chi_{;\nu})+\widetilde{\mathcal{E}}(\Box\chi)_{;\nu}\nn\\
&&+\,\widetilde{\mathcal{A}}\,\myprime\,\psi^{;\mu}\chi_{;\mu}\chi_{;\nu}
+\,\widetilde{\mathcal{B}}\,\myprime\,\chi^{;\mu}\chi_{;\mu}\psi_{;\nu}
+\,\widetilde{\mathcal{E}}\,\myprime\,\Box\chi\,\psi_{;\nu}
+\,\widetilde{\mathcal{C}}\,\myprime \psi^{;\mu}\chi_{;\mu;\nu}\,.
\label{wgksk}
\end{eqnarray}
Since $\Box(\psi_{;\nu})=(\Box\psi)_{;\nu}+R_{\mu\nu}\psi^{;\mu}$,
$\Box(\chi_{;\nu})=(\Box\chi)_{;\nu}+R_{\mu\nu}\chi^{;\mu}$, the expressions (\ref{wksh}),
(\ref{wgksk}) become
\begin{eqnarray}
\vartheta_{\mu\nu}^{\,\,\,\,\,\,\,;\mu}&=&
(\mathcal{A}'\!+\!\mathcal{B}\hspace{0.15mm}'\,)\psi^{;\mu}\psi_{;\mu}\psi_{;\nu}
+(\mathcal{E}\hspace{0.1mm}'\!+\!\mathcal{A})\Box\psi\,\psi_{;\nu}
+(\mathcal{C}\hspace{0.1mm}'\!+\!\mathcal{A}\!+\!2\mathcal{B})
\psi^{;\mu}\psi_{;\mu;\nu}+(\mathcal{C}+\mathcal{E})\,(\Box\psi)_{;\nu}
+\mathcal{C}R_{\mu\nu}\psi^{;\mu}\nn\\
&&+\,\,\,\dispdot[.1ex]{\!\!\mathcal{A}}\,\psi^{;\mu}\chi_{;\mu}\psi_{;\nu}
+\,\,\dispdot[.1ex]{\!\!\mathcal{B}}\,\psi^{;\mu}\psi_{;\mu}\chi_{;\nu}
+\,\,\dispdot[.1ex]{\!\!\mathcal{E}}\,\Box\psi\,\chi_{;\nu}
+\,\,\dispdot[.1ex]{\!\!\mathcal{C}}\,\chi^{;\mu}\psi_{;\mu;\nu}\,,
\label{wjks}
\end{eqnarray}
\begin{eqnarray}
\widetilde{\vartheta}_{\mu\nu}^{\,\,\,\,\,\,\,;\mu}&=&
(\,\,\dispdot[.1ex]{\!\!\widetilde{\mathcal{A}}}\!+\,
\dot{\!\!\widetilde{\mathcal{B}}}\,)\chi^{;\mu}\chi_{;\mu}\chi_{;\nu}
+(\,\,\dot{\!\!\widetilde{\mathcal{E}}}\!+\!\widetilde{\mathcal{A}}\,)\Box\chi\,\chi_{;\nu}
+(\,\,\dot{\!\!\widetilde{\mathcal{C}}}\!+\!
\widetilde{\mathcal{A}}\!+\!2\widetilde{\mathcal{B}}\,)\chi^{;\mu}\chi_{;\mu;\nu}
+(\widetilde{\mathcal{C}}+\widetilde{\mathcal{E}})\,(\Box\chi)_{;\nu}
+\widetilde{\mathcal{C}}R_{\mu\nu}\chi^{;\mu}\nn\\
&&+\,\widetilde{\mathcal{A}}\,\myprime\,\psi^{;\mu}\chi_{;\mu}\chi_{;\nu}
+\,\widetilde{\mathcal{B}}\,\myprime\,\chi^{;\mu}\chi_{;\mu}\psi_{;\nu}
+\,\widetilde{\mathcal{E}}\,\myprime\,\Box\chi\,\psi_{;\nu}
+\,\widetilde{\mathcal{C}}\,\myprime \psi^{;\mu}\chi_{;\mu;\nu}\,.
\label{wgyks}
\end{eqnarray}
Finally, using (\ref{hshk}), equations (\ref{wjks}), (\ref{wgyks}) become
\begin{eqnarray}
\vartheta_{\mu\nu}^{\,\,\,\,\,\,\,;\mu}&\!\!\!\!=\!\!\!\!&
\Big(\mathcal{A}'\!+\!\mathcal{B}'\!+\!\frac{1}{2}\mathcal{A}\mathcal{C}\!-\!
\mathcal{B}\mathcal{C}\Big)\psi^{;\mu}\psi_{;\mu}\psi_{;\nu}
\!-\!\big(\mathcal{C}\mathcal{H}\!-\,\dispdot[.1ex]{\!\!\mathcal{A}}\big)
\psi^{;\mu}\chi_{;\mu}\psi_{;\nu}
\!-\!\Big(\mathcal{C}\widetilde{\mathcal{B}}+\frac{1}{2}\mathcal{C}\widetilde{\mathcal{A}}\Big)
\chi^{;\mu}\chi_{;\mu}\psi_{;\nu}
+\mathcal{C}\widetilde{\mathcal{A}}\,\psi^{;\mu}\chi_{;\mu}\chi_{;\nu}\nn\\
&&+\big(\mathcal{C}\mathcal{F}\!+\,\dispdot[.1ex]{\!\!\mathcal{B}}\big)
\psi^{;\mu}\psi_{;\mu}\chi_{;\nu}
\!+\!\Big(\mathcal{E}\hspace{0.2mm}'\!+\!\mathcal{A}\!-\!\mathcal{C}\mathcal{E}\!-\!\frac{1}{2}
\mathcal{C}^{2}\Big)\Box\psi\,\psi_{;\nu}
\!-\!\Big(\mathcal{C}\widetilde{\mathcal{E}}+\frac{1}{2}\mathcal{C}\widetilde{\mathcal{C}}\Big)
\Box{\chi}\,\psi_{;\nu}
+\,\,\dispdot[.1ex]{\!\!\mathcal{E}}\,\Box\psi\,\chi_{;\nu}
+(\mathcal{C}\!+\!\mathcal{E})(\Box\psi)_{;\nu}\nn\\
&&\,+(\mathcal{C}\hspace{0.1mm}'\!+\!\mathcal{A}\!+\!2\mathcal{B}\!+\!\mathcal{C}^{2})
\psi^{;\mu}\psi_{;\mu;\nu}
+\,\,\dispdot[.1ex]{\!\!\mathcal{C}}\,\chi^{;\mu}\psi_{;\mu;\nu}
+\mathcal{C}\widetilde{\mathcal{C}}\,\psi^{;\mu}\chi_{;\mu;\nu}
+\bar{\Lambda}\mathcal{C}e^{\psi}\psi_{;\nu}+8\pi \bar{G}\mathcal{C} e^{\chi}
\Big(T_{\mu\nu}\!-\!\frac{1}{2}Tg_{\mu\nu}\!\Big)\psi^{;\mu}\,,
\label{sjkj}
\end{eqnarray}
\begin{eqnarray}
\widetilde{\vartheta}_{\mu\nu}^{\,\,\,\,\,\,\,;\mu}&\!\!\!\!=\!\!\!\!&
\Big(\,\dispdot[.1ex]{\!\!\widetilde{\mathcal{A}}}\!+\,
\dot{\!\!\widetilde{\mathcal{B}}}\!+\!\frac{1}{2}\widetilde{\mathcal{A}}\,\widetilde{\mathcal{C}}
\!-\!\widetilde{\mathcal{B}}\,\widetilde{\mathcal{C}}\,\Big)\chi^{;\mu}\chi_{;\mu}\chi_{;\nu}
\!-\!\big(\mathcal{\mathcal{H}}\widetilde{\mathcal{C}}
\!-\!\widetilde{\mathcal{A}}\hspace{0.2mm}\myprime\big)\,\psi^{;\mu}\chi_{;\mu}\chi_{;\nu}
\!-\!\Big(\mathcal{B}\widetilde{\mathcal{C}}+\frac{1}{2}\mathcal{A}\widetilde{\mathcal{C}}\Big)
\psi^{;\mu}\psi_{;\mu}\chi_{;\nu}
+\mathcal{A}\widetilde{\mathcal{C}}\,\psi^{;\mu}\chi_{;\mu}\psi_{;\nu}\nn\\
&&+\big(\mathcal{\mathcal{F}}\widetilde{\mathcal{C}}+\!\widetilde{\mathcal{B}}\,\myprime\big)
\,\chi^{;\mu}\chi_{;\mu}\psi_{;\nu}
+\Big(\,\dispdot[.1ex]{\!\!\widetilde{\mathcal{E}}}
\!+\!\widetilde{\mathcal{A}}\!-\!\widetilde{\mathcal{C}}\,\widetilde{\mathcal{E}}
\!-\!\frac{1}{2}\widetilde{\mathcal{C}}^{\,2}\Big)\Box\chi\,\chi_{;\nu}
\!-\!\Big(\mathcal{E}\widetilde{\mathcal{C}}+\frac{1}{2}\mathcal{C}\widetilde{\mathcal{C}}\Big)
\Box{\psi}\,\chi_{;\nu}
+\widetilde{\mathcal{E}}\hspace{0.25mm}\myprime\,\Box\chi\,\psi_{;\nu}
+(\widetilde{\mathcal{C}}\!+\!\widetilde{\mathcal{E}})(\Box\chi)_{;\nu}\nn\\
&&+\big(\,\,\dispdot[.1ex]{\!\!\widetilde{\mathcal{C}}}+\!\widetilde{\mathcal{A}}
\!+\!2\widetilde{\mathcal{B}}\!+\!\widetilde{\mathcal{C}}^{\,2}\big)\chi^{;\mu}\chi_{;\mu;\nu}
+\widetilde{\mathcal{C}}\,\myprime \psi^{;\mu}\chi_{;\mu;\nu}
+\mathcal{C}\widetilde{\mathcal{C}}\,\chi^{;\mu}\psi_{;\mu;\nu}
+\bar{\Lambda}\widetilde{\mathcal{C}}e^{\psi}\chi_{;\nu}
+8\pi \bar{G}\widetilde{\mathcal{C}} e^{\chi}
\Big(T_{\mu\nu}\!-\!\frac{1}{2}Tg_{\mu\nu}\!\Big)\chi^{;\mu}\,.
\label{sjhy}
\end{eqnarray}

All in all, the consistency condition (\ref{swsbn}) becomes
\begin{eqnarray}
&&\!\!\!\!\!\Big(\!\mathcal{A}'\!+\!\mathcal{B}\hspace{0.3mm}'\!+\!\frac{1}{2}\mathcal{A}\mathcal{C}
\!-\!\mathcal{B}\mathcal{C}\Big)\psi^{;\mu}\psi_{;\mu}\psi_{;\nu}
\!+\!\big(\mathcal{F}\hspace{0.3mm}'\!+\!\mathcal{H}\hspace{0.2mm}'\!+\!\mathcal{A}
\widetilde{\mathcal{C}}\!-\!\mathcal{C}\mathcal{H}\!+\,\dispdot[.1ex]{\!\!\mathcal{A}}\big)
\psi^{;\mu}\chi_{;\mu}\psi_{;\nu}
\!+\!\Big(\,\,\dispdot[.1ex]{\!\!\mathcal{F}}+\mathcal{F}\widetilde{\mathcal{C}}
\!-\!\mathcal{C}\widetilde{\mathcal{B}}\!-\!\frac{1}{2}\mathcal{C}\widetilde{\mathcal{A}}
+\!\widetilde{\mathcal{B}}\,\myprime\Big)
\chi^{;\mu}\chi_{;\mu}\psi_{;\nu}\nn\\
&&\!\!\!\!\!+\big(\,\,\dispdot[.1ex]{\!\!\mathcal{F}}\!+\!\,\,\dispdot[.1ex]{\!\!\mathcal{H}}
\!+\!\mathcal{C}\widetilde{\mathcal{A}}\!-\!\mathcal{H}\widetilde{\mathcal{C}}
+\!\widetilde{\mathcal{A}}\hspace{0.2mm}\myprime\big)\psi^{;\mu}\chi_{;\mu}\chi_{;\nu}
\!+\!\Big(\!\mathcal{\mathcal{F}}\hspace{0.2mm}'\!+\!\mathcal{C}\mathcal{F}\!-\!\mathcal{B}
\widetilde{\mathcal{C}}
\!-\!\frac{1}{2}\mathcal{A}\widetilde{\mathcal{C}}+\,\dispdot[.1ex]{\!\!\mathcal{B}}\Big)
\psi^{;\mu}\psi_{;\mu}\chi_{;\nu}
\!+\!\Big(\,\dispdot[.1ex]{\!\!\widetilde{\mathcal{A}}}+\,\dispdot[.1ex]{\!\!\widetilde{\mathcal{B}}}
+\!\frac{1}{2}\widetilde{\mathcal{A}}\,\widetilde{\mathcal{C}}\!-\!\widetilde{\mathcal{B}}\,
\widetilde{\mathcal{C}}\Big)\chi^{;\mu}\chi_{;\mu}\chi_{;\nu}\nn\\
&&\!\!\!\!\!
+\Big(\!\mathcal{E}'\!+\!\mathcal{A}\!-\!\mathcal{C}\mathcal{E}\!-\!\frac{1}{2}\mathcal{C}^{2}\Big)
\Box\psi\,\psi_{;\nu}
\!+\!\Big(\!\mathcal{F}\!-\!\mathcal{C}\widetilde{\mathcal{E}}
\!-\!\frac{1}{2}\mathcal{C}\widetilde{\mathcal{C}}+\!\widetilde{\mathcal{E}}\hspace{0.25mm}\myprime\Big)
\Box{\chi}\,\psi_{;\nu}
\!+\!\Big(\!\mathcal{F}\!-\!\mathcal{E}\widetilde{\mathcal{C}}
\!-\!\frac{1}{2}\mathcal{C}\widetilde{\mathcal{C}}+\,\dispdot[.1ex]{\!\!\mathcal{E}}\Big)\Box{\psi}
\,\chi_{;\nu}
\!+\!\Big(\,\dispdot[.1ex]{\!\!\widetilde{\mathcal{E}}}\!+\!\widetilde{\mathcal{A}}\!-\!
\widetilde{\mathcal{C}}\widetilde{\mathcal{E}}\!-\!
\frac{1}{2}\widetilde{\mathcal{C}}^{\,2}\Big)\Box\chi\,\chi_{;\nu}\nn\\
&&\!\!\!\!\!+(\mathcal{C}\!+\!\mathcal{E})(\Box\psi)_{;\nu}
\!+\!(\widetilde{\mathcal{C}}\!+\!\widetilde{\mathcal{E}})(\Box\chi)_{;\nu}
\!+\!\big(\mathcal{C}\hspace{0.2mm}'\!+\!\mathcal{A}\!+\!2\mathcal{B}\!+\!\mathcal{C}^{2}\big)
\psi^{;\mu}\psi_{;\mu;\nu}
\!+\!\big(\mathcal{C}\widetilde{\mathcal{C}}\!+\!\mathcal{F}\!+\!\mathcal{H}
+\,\,\dispdot[.1ex]{\!\!\mathcal{C}}\big)\chi^{;\mu}\psi_{;\mu;\nu}
\!+\!\big(\mathcal{C}\widetilde{\mathcal{C}}\!+\!\mathcal{F}\!+\!\mathcal{H}
+\widetilde{\mathcal{C}}\,\myprime\big)\psi^{;\mu}\chi_{;\mu;\nu}\nn\\
&&\!\!\!\!\!+\big(\,\,\dispdot[.1ex]{\!\!\widetilde{\mathcal{C}}}\!+\!\widetilde{\mathcal{A}}
\!+\!2\widetilde{\mathcal{B}}\!+\!\widetilde{\mathcal{C}}^{\,2}\big)\chi^{;\mu}
\chi_{;\mu;\nu}\!+\!\bar{\Lambda}(\mathcal{C}\!-\!1)e^{\psi}\psi_{;\nu}
\!+\!\bar{\Lambda}\widetilde{\mathcal{C}}e^{\psi}\chi_{;\nu}
\!+\!8\pi\Big[\!\big(GT_{\mu\nu}\big)^{;\mu}\!+\!G\Big(\!T_{\mu\nu}\!-\!\frac{1}{2}Tg_{\mu\nu}\!\Big)
\big(\mathcal{C}\psi^{;\mu}\!+\!\widetilde{\mathcal{C}}\chi^{;\mu}\big)\!\Big]\!=\!0\,.\label{lokyu}
\end{eqnarray}
\indent The first thing to say about equation (\ref{lokyu}), as mentioned above, is that in our
treatment the functions $\Lambda(x),G(x)$, i.e. the functions $\psi(x),\chi(x)$, are considered
as known spacetime functions. In this sense, equation (\ref{lokyu}) forms in general a set of
four equations containing the ten unknown coefficients $\mathcal{A}, \mathcal{B},
\mathcal{C}, \mathcal{E}, \widetilde{\mathcal{A}}, \widetilde{\mathcal{B}}, \widetilde{\mathcal{C}},
\widetilde{\mathcal{E}}, \mathcal{F}, \mathcal{H}$ and the energy-momentum tensor $T_{\mu\nu}$.
Equation (\ref{lokyu}) cannot be considered as the equation of motion for $T_{\mu\nu}$
because then, various of these coefficients (or maybe all, depending on the underlying spacetime
symmetry) would remain completely undetermined and this situation would be ill-defined.
If, on the other hand, $T_{\mu\nu}$ is given independently by an add-hoc conservation equation
(e.g. $T_{\mu\nu}^{\,\,\,\,\,\,;\mu}=0$), then $T_{\mu\nu}$ is also considered to be known in
(\ref{lokyu}), as $\psi(x),\chi(x)$. In this case, (\ref{lokyu}) would form a set of four
(in general) partial differential equations for the ten coefficient functions of $\psi,\chi$ and
it is obvious again that various of these coefficients would remain undetermined.

One could actually try to evade this issue of enumeration by adopting that in the initial equations
(\ref{njgc}) there are only present the four coefficients $\mathcal{A}, \mathcal{B}, \mathcal{C},
\mathcal{E}$ (or analogously, e.g. the four coefficients
$\widetilde{\mathcal{A}}, \widetilde{\mathcal{B}},
\widetilde{\mathcal{C}}, \widetilde{\mathcal{E}}$) and the rest of the kinetic or mixed terms are
absent. In this case, the enumeration of equations and unknowns in principle agrees (supposed that
$T_{\mu\nu}$ obeys a pre-assumed equation of motion) and (\ref{lokyu}) gets a much simpler form with
fewer terms. Then, the coefficients $\mathcal{A}, \mathcal{B}, \mathcal{C}, \mathcal{E}$,
which are under quest, could be either assumed to be functions of $\psi$ or that they are just
spacetime functions. If the coefficients are functions of $\psi$, it is clear that (\ref{lokyu}) is
inconsistent since (\ref{lokyu}) contains $x^{\mu}$ explicitly through the various kinetic terms
(e.g. $\psi^{;\mu}\psi_{;\mu}\psi_{;\nu}$), and therefore, solving the ordinary differential
equations (\ref{lokyu}) for the coefficients, only incidentally would they turn out to be indeed
functions of $\psi$ without containing $x^{\mu}$; the simple and manageable situation where just
a single $x^{\mu}$ appears in the equations, is certainly special. If the coefficients are just
spacetime functions, in (\ref{lokyu}) the derivative terms
$\big(\mathcal{A}'\!+\!\mathcal{B}\hspace{0.3mm}'\big)\psi^{;\mu}\psi_{;\mu}\psi_{;\nu}$,
$\mathcal{E}\hspace{0.2mm}'\Box\psi\,\psi_{;\nu}$,
$\mathcal{C}\hspace{0.2mm}'\psi^{;\mu}\psi_{;\mu;\nu}$ are replaced by
$\big(\mathcal{A}^{;\mu}+\mathcal{B}^{;\mu}\big)\psi_{;\mu}\psi_{;\nu}$,
$\mathcal{E}_{;\nu}\Box\psi$, $\mathcal{C}^{;\mu}\psi_{;\mu;\nu}$ respectively. Then, the
coefficients $\mathcal{A}(x), \mathcal{B}(x), \mathcal{C}(x), \mathcal{E}(x)$ could in principle
be derived by solving the partial differential equations (\ref{lokyu}). However, this could be done
only case by case, i.e. for a specific symmetry of the geometrical configuration, along with a
specific choice of $\psi(x),\chi(x)$. It is not evident if this could even be called ``a theory''
since by passing e.g. from cosmology to spherically symmetric configurations, the solving process
of (\ref{lokyu}) should be done anew, finding the new coefficients and the new equations (\ref{njgc}),
which would be valid only for this particular background. Moreover, by solving such partial
differential equations (or maybe ordinary differential equations if the configuration is of high
symmetry), integration constants or integration functions should appear in the solution for the
coefficients. Therefore, e.g. the arising ``theory'' of cosmology could have a different number of
integration constants than the arising ``theory'' of black holes, let apart the more complicated
situations where integration functions appear. These integration constants, which are not the
familiar spacetime integration constants, appear practically as new parameters in the equations.
What is peculiar and lacks physical significance is the inability to determine these new parameters
by performing measurements and experiments in different regimes of the real cosmos (since the
parameters depend on the specific geometry configuration), something
which is standard in all successful physical theories. Of course, the appearance of integration
functions, which is unavoidable in configurations with less spacetime symmetries, does not have the
interpretation of new parameters and the arbitrariness entering by such functions makes the total
picture ill-defined and the ``theory'' even more pointless.

The result of the above discussion is that the only physically reasonable way to treat equations
(\ref{lokyu}) is to try to find well-defined solutions for all coefficients, which at the same time
are global (i.e. independent of a specific spacetime configuration), and therefore indeed form a
physical theory. This will be the case in the following, with the extra merit that the arising
coefficients will turn out to be unique pure numbers, they will not contain integration constants or
integration functions. Even if hypothetically such integration constants or functions appeared (which
will not be the case here), maybe by including fourth derivatives in the equations of motions, this
situation would be acceptable (particularly for integration constants
playing the role of new parameters) and quite different than the discussion about integration
constants or functions given a few lines above, because now these integration constants (or
functions) would be global and the same for any configuration, thus participating in a well-defined
physical theory.

We finally come to the point to explain how one finds the coefficients $\mathcal{A},
\widetilde{\mathcal{A}}, ...$ from (\ref{lokyu}) and constructs the theory. The recipe is that each
prefactor of any kinetic/derivative term in (\ref{lokyu}) (e.g. $\psi^{;\mu}\psi_{;\mu}\psi_{;\nu}$,
$(\Box\chi)_{;\nu}$, $\psi_{;\nu}$, etc.) should vanish. All such kinetic terms in (\ref{lokyu})
can be considered as functionally independent in the sense that there is no way to convert one such
term to some others. This means that the satisfaction of (\ref{lokyu}) occurs identically, i.e. for
all $\psi(x),\chi(x)$. This does not mean that we are interested in all possible $\psi(x),\chi(x)$
functions since these functions are supposed to be uniquely determined in the context of the theory
at hand, either this is the Asymptotic Safety scenario of quantum gravity (where $\Lambda(k),G(k)$
are the output of the RG flow equations and a relation between the energy scale $k$ and a spacetime
scale should somehow be introduced) or some other gravity theory with varying constants. This identical
satisfaction of (\ref{lokyu}) basically means that this is the only way to assure that the arising
coefficients are indeed independent of the spacetime configuration and therefore are capable to form a
well-defined physical theory{\footnote{It is now clear that the inclusion in (\ref{njgc}) of kinetic
terms with curvature couplings, such as $GR_{\kappa(\mu}\psi_{;\nu)}\psi^{;\kappa}$, would lead
through the Bianchi identities to the inclusion of terms containing a derivative of $R_{\mu\nu}$
in (\ref{lokyu}). Such a derivative term cannot be canceled by any other term and cannot be
substituted by the new equation (\ref{hshk}). Therefore, such kinetic terms with curvature couplings
cannot be included in the equations of motion (\ref{njgc}).}}.
As also realized from the previous discussion, in this vanishing of
the prefactors of kinetic terms, terms in (\ref{lokyu}) containing the energy-momentum tensor
$T_{\mu\nu}$ cannot participate since then, the arising coefficients would become background
dependent (and also energy-momentum dependent). As an additional result of the vanishing of the
coefficients, the conservation equation of $T_{\mu\nu}$ also arises uniquely from (\ref{lokyu})
by vanishing the combination of all terms containing $T_{\mu\nu}$. Hence, we arrive at the following
set of equations
\begin{eqnarray}
&&\!\mathcal{A}'\!+\!\mathcal{B}\hspace{0.3mm}'\!+\!\frac{1}{2}\mathcal{A}\mathcal{C}
\!-\!\mathcal{B}\mathcal{C}=0\,\,\,\,\,\,\,\,\,,\,\,\,\,\,\,\,\,\,
\mathcal{F}\hspace{0.3mm}'\!+\!\mathcal{H}\hspace{0.2mm}'\!+\!\mathcal{A}
\widetilde{\mathcal{C}}\!-\!\mathcal{C}\mathcal{H}\!+\,\dispdot[.1ex]{\!\!\mathcal{A}}=0
\,\,\,\,\,\,\,\,\,,\,\,\,\,\,\,\,\,\,
\mathcal{\mathcal{F}}\hspace{0.2mm}'\!+\!\mathcal{C}\mathcal{F}\!-\!\mathcal{B}
\widetilde{\mathcal{C}}\!-\!
\frac{1}{2}\mathcal{A}\widetilde{\mathcal{C}}+\,\dispdot[.1ex]{\!\!\mathcal{B}}=0\,,\label{gtey}\\
&&\!\mathcal{E}'\!+\!\mathcal{A}\!-\!\mathcal{C}\mathcal{E}\!-\!\frac{1}{2}\mathcal{C}^{2}=0
\,\,\,\,\,\,\,\,\,,\,\,\,\,\,\,\,\,\,
\mathcal{C}\hspace{0.2mm}'\!+\!\mathcal{A}\!+\!2\mathcal{B}\!+\!\mathcal{C}^{2}=0\,,\\
&&\,\dispdot[.1ex]{\!\!\mathcal{F}}+\mathcal{F}\widetilde{\mathcal{C}}
\!-\!\mathcal{C}\widetilde{\mathcal{B}}\!-\!\frac{1}{2}\mathcal{C}\widetilde{\mathcal{A}}
+\!\widetilde{\mathcal{B}}\,\myprime=0
\,\,\,\,\,\,\,\,\,,\,\,\,\,\,\,\,\,\,
\dispdot[.1ex]{\!\!\mathcal{F}}\!+\!\,\,\dispdot[.1ex]{\!\!\mathcal{H}}
\!+\!\mathcal{C}\widetilde{\mathcal{A}}\!-\!\mathcal{H}\widetilde{\mathcal{C}}
+\!\widetilde{\mathcal{A}}\hspace{0.2mm}\myprime=0\,,\\
&&\,\dispdot[.1ex]{\!\!\widetilde{\mathcal{A}}}+\,\dispdot[.1ex]{\!\!\widetilde{\mathcal{B}}}
+\!\frac{1}{2}\widetilde{\mathcal{A}}\,\widetilde{\mathcal{C}}\!-\!\widetilde{\mathcal{B}}\,
\widetilde{\mathcal{C}}=0\,\,\,\,\,\,\,\,\,,\,\,\,\,\,\,\,\,\,
\dispdot[.1ex]{\!\!\widetilde{\mathcal{E}}}\!+\!\widetilde{\mathcal{A}}\!-\!
\widetilde{\mathcal{C}}\widetilde{\mathcal{E}}\!-\!
\frac{1}{2}\widetilde{\mathcal{C}}^{\,2}=0\,\,\,\,\,\,\,\,\,,\,\,\,\,\,\,\,\,\,
\dispdot[.1ex]{\!\!\widetilde{\mathcal{C}}}\!+\!\widetilde{\mathcal{A}}
\!+\!2\widetilde{\mathcal{B}}\!+\!\widetilde{\mathcal{C}}^{\,2}=0\,,\\
&&\!\mathcal{F}\!-\!\mathcal{C}\widetilde{\mathcal{E}}
\!-\!\frac{1}{2}\mathcal{C}\widetilde{\mathcal{C}}+\!\widetilde{\mathcal{E}}\hspace{0.25mm}\myprime=0
\,\,\,\,\,\,\,\,\,,\,\,\,\,\,\,\,\,\,
\mathcal{F}\!-\!\mathcal{E}\widetilde{\mathcal{C}}
\!-\!\frac{1}{2}\mathcal{C}\widetilde{\mathcal{C}}+\,\dispdot[.1ex]{\!\!\mathcal{E}}=0
\,\,\,\,\,\,\,\,\,,\,\,\,\,\,\,\,\,\,
\mathcal{C}\!+\!\mathcal{E}=0\,\,\,\,\,\,\,\,\,,\,\,\,\,\,\,\,\,\,
\widetilde{\mathcal{C}}\!+\!\widetilde{\mathcal{E}}=0\,,\\
&&\!\mathcal{C}\widetilde{\mathcal{C}}\!+\!\mathcal{F}\!+\!\mathcal{H}
+\,\,\dispdot[.1ex]{\!\!\mathcal{C}}=0\,\,\,\,\,\,\,\,\,,\,\,\,\,\,\,\,\,\,
\mathcal{C}\widetilde{\mathcal{C}}\!+\!\mathcal{F}\!+\!\mathcal{H}
+\widetilde{\mathcal{C}}\,\myprime=0\,\,\,\,\,\,\,\,\,,\,\,\,\,\,\,\,\,\,
\mathcal{C}=1\,\,\,\,\,\,\,\,\,,\,\,\,\,\,\,\,\,\,
\widetilde{\mathcal{C}}=0\,,\label{gteyn}
\end{eqnarray}
\begin{eqnarray}
&&\!\!\!\!\!\!\!\!\!\!\!\!\!\!\!\!\!\!\!\!\!\!\!\!\!\!\!\!\!\!\!\!\!\!\!\!\!\!\!\!\!\!
\!\!\!\!\!\!\!\!\!\!\!\!\!\!\!\!\!\!\!\!\!\!\!\!\!\!\!\!\!\!\!\!\!\!\!\!\!\!\!\!\!\!
\!\!\!\!\!\!\!\!\!\!\!\!\!\!\!\!\!\!\!\!\!\!\!\!\!\!\!\!\!\!\!\!\!\!\!\!\!\!\!\!\!\!
\big(GT_{\mu\nu}\big)^{;\mu}\!+\!G\Big(\!T_{\mu\nu}\!-\!\frac{1}{2}Tg_{\mu\nu}\!\Big)
\big(\mathcal{C}\psi^{;\mu}\!+\!\widetilde{\mathcal{C}}\chi^{;\mu}\big)=0\,.\label{khywy}
\end{eqnarray}
Equations (\ref{gtey})-(\ref{gteyn}) form a system of eighteen equations, most of them
partial differential equations, for the ten coefficient functions $\mathcal{A},
\widetilde{\mathcal{A}}, ...$ of $\psi,\chi$ in search. However, the solution of this system turns
out to be only algebraic and unique, given by
\begin{equation}
\mathcal{A}=-\frac{1}{2}\,\,\,\,\,,\,\,\,\,\,\mathcal{B}=-\frac{1}{4}\,\,\,\,\,,\,\,\,\,\,
\mathcal{C}=1\,\,\,\,\,,\,\,\,\,\,\mathcal{E}=-1\,\,\,\,\,,\,\,\,\,\,
\widetilde{\mathcal{A}}\,=\,\widetilde{\mathcal{B}}\,=\,\widetilde{\mathcal{C}}\,=\,
\widetilde{\mathcal{E}}\,=\,\mathcal{H}=\mathcal{F}=0\,.
\end{equation}

The gravitational equations (\ref{njgc}) finally become
\begin{equation}
\boxed{G_{\mu\nu}=-\bar{\Lambda}\,e^{\psi} g_{\mu\nu}
-\frac{1}{2}\psi_{;\mu}\psi_{;\nu}-\frac{1}{4}g_{\mu\nu}\psi^{;\rho}
\psi_{;\rho}+\psi_{;\mu;\nu}-g_{\mu\nu}\Box\psi+8\pi G T_{\mu\nu}}
\label{skrvgd}
\end{equation}
and the conservation equation (\ref{khywy}) gets the form
\begin{equation}
\boxed{\big(GT_{\mu\nu}\big)^{;\mu}\!+\!G\Big(T_{\mu\nu}\!-\!\frac{1}{2}Tg_{\mu\nu}\Big)
\psi^{;\mu}=0}\,.
\label{skbmn}
\end{equation}
{\textit{Equations (\ref{skrvgd}), (\ref{skbmn}) form the ultimate result of our search for a consistent
gravitational theory with varying $\Lambda=\bar{\Lambda}\,e^{\psi}$ and $G=\bar{G}\,e^{\chi}$}}.
The conservation equation (\ref{skbmn}) can be considered as the consistency condition which arises
through Bianchi identities from the modified Einstein equations (\ref{skrvgd}). The way of derivation
shows that the system (\ref{skrvgd}), (\ref{skbmn}) is the unique, mathematically and physically
consistent, modification of Einstein gravity with varying $\Lambda,G$, and containing up to
second derivatives in $g_{\mu\nu},\Lambda,G$. Obviously, constant $\Lambda$ and $G$ reduce equations
(\ref{skrvgd}), (\ref{skbmn}) to the Einstein equations with a cosmological constant and the standard
conservation of matter.

In the case where $G=G(\Box)$, all equations (\ref{lokyu}), (\ref{gtey})-(\ref{khywy}) remain intact
(of course with the non-trivial terms and equations remaining when derivatives of $\chi$ are neglected),
and finally the system (\ref{skrvgd}), (\ref{skbmn}) is exactly the same. Note that in (\ref{skbmn}),
$G(\Box)$ acts on the $T_{\mu\nu}$ components and not on $\psi^{;\mu}$.

We note that there are only $\Lambda$-kinetic terms but no $G$-kinetic terms in the gravitational
equation (\ref{skrvgd}). It seems that the explanation of this comes from the second and third
term of the last line of equation (\ref{lokyu}). The coefficient $\mathcal{C}$ of
$\vartheta_{\mu\nu}$ acquires a non-vanishing value because of the presence of the cosmological
constant term in the equations, so the covariant derivative of $\vartheta_{\mu\nu}$ cancels against
$\Lambda$. However, there is no such a corresponding term for $G$ in the equations (\ref{njgc}) and
the coefficient $\widetilde{\mathcal{C}}$ of $\widetilde{\vartheta}_{\mu\nu}$ vanishes since the
cancelation of the covariant derivative of $\widetilde{\vartheta}_{\mu\nu}$ occurs identically, that
is against zero. Since there are no $G$-kinetic terms in (\ref{skrvgd}), in vacuum (absence of
matter) there is no impact from a possible variability of $G$; only the presence of matter reveals
a spacetime-dependent $G$ (similarly to the situation with a constant Newton's $G$, which is
meaningful and measurable only in the presence of matter). So, in vacuum, the Bianchi identities from
(\ref{skrvgd}) are identically satisfied for any $\Lambda(x)$, in agreement with
\cite{Kofinas:2015sna}. In the conservation equation (\ref{skbmn}), we see that there are still
interaction terms with derivatives of $\Lambda$. Moreover, in (\ref{skbmn}) derivative terms of $G$
also appear, however only through the combination $GT_{\mu\nu}$, while if $\widetilde{\mathcal{C}}$
was non-zero, extra $G^{;\mu}$ interaction terms would appear in (\ref{skbmn}) as (\ref{khywy}) shows.

Equations (\ref{skrvgd}), (\ref{skbmn}) do not cover the situation where the cosmological
constant is vanishing ($\bar{\Lambda}=0$), but still some $\psi$-kinetic terms exist in
(\ref{njgc}) which might have an independent physical meaning. Indeed, in this case
the second and third terms of the last line of equation (\ref{lokyu}) are absent and so $\mathcal{C},
\widetilde{\mathcal{C}}$ do not take the values one and zero respectively as here.

Asymptotically Safe gravity is a theory with varying $G$ and $\Lambda$, where equations
(\ref{skrvgd}), (\ref{skbmn}) can be applied. In the case of AS gravity, there is another approach
which is not based on field equations such as the improved field equations (\ref{skrvgd}),
(\ref{skbmn}), but the quantum corrections of $G$ and $\Lambda$ enter through an
improved action by variation with respect to $g_{\mu\nu}$. In this approach the gravitational
``constant'' plays the role of Lagrange multiplier in the gravitational action even in vacuum and
the variation of this action provides derivative terms of $G$ in the gravitational equations of motion
\cite{Reuter:2003ca} (however, since the derivation scheme is somewhat implicit, such consistent
equations are not yet known in general). In our approach the consistent gravitational equation
(\ref{skrvgd}), where the $G$-kinetic terms are absent, implies that although the action is certainly
a necessary devise for obtaining the $\beta$-functions of the RG flow for the various couplings,
however the equations of macroscopic phenomenological gravity may be derivable directly at the level
of the equations of motion taking into account their consistency through the Bianchi identities.

\section{General cosmology with time evolving $G$ and $\Lambda$}

In this section we will apply the general equations (\ref{skrvgd}), (\ref{skbmn}) in the case of a
spatially homogeneous and isotropic cosmological metric of the form
\begin{equation}
ds^{2}=-n(t)^{2}dt^{2}+a(t)^{2}\Big[\frac{dr^{2}}{1\!-\!\kappa\,r^{2}}+r^{2}\big(d\theta^{2}
\!+\!\sin^{2}{\!\theta}\,d\phi^{2}\big)\Big]\,, \label{jkwk}
\end{equation}
without specifying the underlying theory or the form of the varying constants $\Lambda$ and $G$.
In (\ref{jkwk}) the function $n(t)$ is the lapse, while the quantity $\kappa=-1,0,1$ (with dimensions
inverse length squared for $\kappa\neq 0$) characterizes the spatial curvature. Since the external
spacetime functions $\Lambda(x),G(x)$ carry the same symmetries, they will be of the form
$\Lambda(t),G(t)$. In Asymptotic Safety in particular, $\Lambda$ and $G$ are given as functions of
the energy scale $k$, i.e. it is $\Lambda(k)$ and $G(k)$, therefore $k$ should be somehow related to
time, either explicitly or implicitly through some cosmological parameter. We consider a diagonal
energy-momentum tensor $T^{\mu}_{\nu}$, so we take as matter content a fluid with energy density
$\rho$ and pressure $P$. At this stage we do not restrict or specify the form of the pressure $P$;
it can contain the standard thermodynamic pressure with an arbitrary equation of state or also
possible non-equilibrium parts. Due to the working symmetries of isotropy and homogeneity,
off-diagonal contributions from shear viscosity and energy fluxes are disregarded. The energy
momentum tensor is
\begin{equation}
T^{\mu\nu}=(\rho+P) u^{\mu}u^{\nu}+P g^{\mu\nu}\,,
\label{tmn}
\end{equation}
with $u^{\mu}$ the fluid 4-velocity.

For the metric (\ref{jkwk}) the non-vanishing components of the Einstein tensor $G_{\nu}^{\mu}$ are
\begin{eqnarray}
G^{t}_{t}&=&-3\Big(\!H^{2}+\frac{\kappa}{a^{2}}\Big)\,, \nn \\
G^{i}_{j}&=&-\Big(\frac{2}{n}\dot{H}+3H^{2}+\frac{\kappa}{a^{2}}\Big)\delta^{i}_{j}\,,
\label{skk}
\end{eqnarray}
where the indices $i,j$ refer to the spatial coordinates, $H=\frac{\dot{a}}{na}$ is the Hubble
parameter, and a dot will denote from now on a differentiation with respect to $t$. It is also
possible to evaluate
\begin{equation}
\label{psi1}
\psi^{;\rho}\psi_{;\rho}=-\frac{\dot{\psi}^{2}}{n^{2}}\,\,\,\,,\,\,\,\,
\Box\psi=-\frac{1}{n}\Big(\frac{\dot{\psi}}{n}\Big)^{^{\!\!\LargerCdot}}-3H\frac{\dot{\psi}}{n}\,,
\end{equation}
the non-vanishing components of $\psi_{;\mu;\nu}$ are
\begin{equation}
\label{comps}
\psi_{;t;t}=n\Big(\frac{\dot{\psi}}{n}\Big)^{^{\!\!\LargerCdot}}\,\,\,\,,\,\,\,\,
\psi_{;r;r}=-\frac{Ha^{2}}{1\!-\!\kappa\,r^{2}}\,\frac{\dot{\psi}}{n}\,\,\,\,,\,\,\,\,
\psi_{;\theta;\theta}=-H
a^{2}r^{2}\frac{\dot{\psi}}{n}\,\,\,\,,\,\,\,\, \psi_{;\phi;\phi}=-H
a^{2}r^{2}\sin^{2}\!\theta\,\frac{\dot{\psi}}{n}\,,
\end{equation}
while the components of the energy-momentum tensor are $T^{t}_{t}=-\rho$, $T^{i}_{j}=P\delta^{i}_{j}$.

Therefore, the two independent components of (\ref{skrvgd}) are
\begin{eqnarray}
\!H^{2}+\frac{\kappa}{a^{2}}\!&=&\!\frac{\bar{\Lambda}}{3}e^{\psi}-H\frac{\dot{\psi}}{n}
-\frac{\dot{\psi}^{2}}{4n^{2}}+\frac{8\pi}{3}G\rho\,,
\label{eqvac1w}\\
\frac{2}{n}\dot{H}+3H^{2}+\frac{\kappa}{a^{2}}
\!&=&\!\bar{\Lambda}e^{\psi}-2H\frac{\dot{\psi}}{n}-\frac{\dot{\psi}^{2}}{4n^{2}}
-\frac{1}{n}\Big(\frac{\dot{\psi}}{n}\Big)^{^{\!\!\LargerCdot}}-8\pi G P\,.
\label{eqvac2w}
\end{eqnarray}
Equation (\ref{eqvac2w}) can be substituted by a combination of (\ref{eqvac1w}), (\ref{eqvac2w}),
namely
\begin{equation}
\frac{1}{n}\dot{H}=\frac{\kappa}{a^{2}}+H\frac{\dot{\psi}}{2n}+\frac{\dot{\psi}^{2}}{4n^{2}}
-\frac{1}{2n}\Big(\frac{\dot{\psi}}{n}\Big)^{^{\!\!\LargerCdot}}-4\pi G (\rho+P)\,.
\label{nhuyes}
\end{equation}
The conservation equation (\ref{skbmn}) gets the form
\begin{equation}
\dot{\rho}+3nH(\rho+P)+\rho\dot{\chi}+\frac{\rho+3P}{2}\dot{\psi}=0\,,
\label{hyeh}
\end{equation}
which implies an energy transfer between the energy density $\rho$ and $G,\Lambda$.

The system of the three equations (\ref{eqvac1w}), (\ref{nhuyes}), (\ref{hyeh}) for the two
unknown functions $a(t),\rho(t)$ is consistent for any $P$, and this is expected since it is the
cosmological reduction of the basic system (\ref{skrvgd}), (\ref{skbmn}). So, one of the equations
(\ref{eqvac1w}), (\ref{nhuyes}), (\ref{hyeh}) is redundant (for any function $P(t)$) and can be
derived from the other two equations, a feature which is reminiscent to the standard FRW cosmology.
Other conservation equations, adopted in \cite{2016PhRvD..94j3514K}, \cite{Zarikas:2018wfv},
induce some constraints on the pressure $P$.

To see the redundancy of (\ref{nhuyes}) explicitly, we differentiate (\ref{eqvac1w}) with respect to
time $t$, we then use (\ref{hyeh}) to eliminate $\dot{\rho}$, and finally use (\ref{eqvac1w}) as it
is once more. The resulting equation is
\begin{equation}
\Big(2H\!+\!\frac{\dot{\psi}}{n}\Big)
\Big[\frac{1}{n}\dot{H}\!-\!\frac{\kappa}{a^{2}}\!-\!H\frac{\dot{\psi}}{2n}
\!-\!\frac{\dot{\psi}^{2}}{4n^{2}}\!+\!\frac{1}{2n}\Big(\frac{\dot{\psi}}{n}\Big)^{^{\!\!\LargerCdot}}
\!+\!4\pi G (\rho+P)\Big]=0\,.
\label{lfej}
\end{equation}
In the case where the prefactor $2H\!+\!\frac{\dot{\psi}}{n}$ of equation (\ref{lfej}) vanishes,
the consistency equation (\ref{lfej}) is satisfied and at the same time equation (\ref{nhuyes})
is not identically satisfied, therefore (\ref{nhuyes}) is not redundant. So, the vanishing of this
prefactor gives $\Lambda=\frac{\bar{\Lambda}c'}{a^{2}}$, where $c'>0$ is integration constant.
Equation (\ref{eqvac1w}) becomes equivalent to $G\rho=\frac{3\kappa-\bar{\Lambda} c'}{8\pi a^{2}}$,
therefore for a positive cosmological constant it should be $\kappa>0$. The conservation equation
(\ref{hyeh}) is integrated to $G\rho=\frac{c''}{a^{2}}$, where $c''>0$ is another integration
constant, therefore $\bar{\Lambda}c'+8\pi c''=3\kappa$. Finally, equation (\ref{nhuyes}) gives
$4\pi G(\rho+P)=\frac{\kappa}{a^{2}}$, and therefore $4\pi c'' (1+P/\rho)=\kappa$,
which provides a constant equation of state for the fluid. As a result, the case we are discussing
with the vanishing of the prefactor has $\Lambda\sim a^{-2}$, $\rho\sim (Ga^{2})^{-1}$ for any $G$
assumed, the equation of state $P=P(\rho)$ is uniquely specified, but there is no equation left to
determine the cosmic evolution itself, i.e. to give $a(t)$, therefore this case is ill-defined.
What remains is that equation (\ref{lfej}) coincides with (\ref{nhuyes}), proving the consistency
of our cosmological system (\ref{eqvac1w}), (\ref{nhuyes}), (\ref{hyeh}).

The most standard assumption is to consider that the fluid is perfect, thus $P$ is solely the
thermodynamic pressure of the fluid. Furthermore, for a constant equation of state parameter
$w=\frac{P}{\rho}$ (general barotropic fluid), the conservation equation (\ref{hyeh}) is integrated
to give
\begin{equation}
G\rho=\frac{c}{a^{3(1+w)}}\,e^{-\frac{1+3w}{2}\psi}\,,
\label{kiyetn}
\end{equation}
or equivalently
\begin{equation}
\Lambda^{\frac{1+3w}{2}}G\rho=\frac{c\,\bar{\Lambda}^{\frac{1+3w}{2}}}{a^{3(1+w)}}\,,
\label{kiyet}
\end{equation}
where $c>0$ is an integration constant (with dimensions inverse length squared) related to the amount
of matter. In (\ref{kiyetn}) or (\ref{kiyet}), any time-dependent functions $G,\psi$ can be assumed.
As a result, equations (\ref{eqvac1w}), (\ref{nhuyes}) become
\begin{eqnarray}
&&H^{2}+\frac{\kappa}{a^{2}}=\frac{\bar{\Lambda}}{3}e^{\psi}-H\frac{\dot{\psi}}{n}
-\frac{\dot{\psi}^{2}}{4n^{2}}+\frac{8\pi c}{3a^{3(1+w)}}\,e^{-\frac{1+3w}{2}\psi}\,,
\label{hywte}\\
&&\frac{1}{n}\dot{H}=\frac{\kappa}{a^{2}}+H\frac{\dot{\psi}}{2n}+\frac{\dot{\psi}^{2}}{4n^{2}}
-\frac{1}{2n}\Big(\frac{\dot{\psi}}{n}\Big)^{^{\!\!\LargerCdot}}
-\frac{4\pi c (1\!+\!w)}{a^{3(1+w)}}\,e^{-\frac{1+3w}{2}\psi}\,,
\label{hwynqo}
\end{eqnarray}
where any time-dependent function $\psi$ can be assumed. Henceforth we will adopt the cosmic time
parameter $t$ with $n=1$.
Equations (\ref{kiyetn}), (\ref{hywte}), (\ref{hwynqo}) are our resulting cosmological equations
which can be applied in any theory with varying gravitational constant $G$ and cosmological
constant $\Lambda=\bar{\Lambda}e^{\psi}$. The consistency of this system is manifested by the
redundancy of one of these three equations, e.g. equation (\ref{hwynqo}).

\section{Early-times cosmology in the theory of Asymptotic Safety}
\label{solutions}

We will now focus on the Asymptotic Safety scenario of quantum gravity in order to describe the
variability of $\Lambda,G$ and implement the previous formulation in a specific and well-defined
theory. In the context of AS scenario, the geometry-independent RG flow
equations predict the running of both $\Lambda$, $G$ in terms of some characteristic
energy scale $k$ describing the physical system, i.e. they provide the functions $\Lambda(k),G(k)$.
However, in order to proceed with the solution of (\ref{hywte}), we have to determine the evolution
of $\psi$ as a function of spacetime, not of $k$. This will come by the selection of a scaling that
associates the energy of RG scale $k$ to a characteristic time or length of the system. This
choice is arbitrary and is not a prediction of Asymptotic Safety framework. In cosmology,
$k$ could be related to e.g. $t,a,H,\rho$, etc., so one can use scalings such as $k\propto 1/t$,
$k\propto H(t)$, etc. The first studies selected this RG scale to be proportional to $1/t$
\cite{Bonanno:2001xi} or $H$ \cite{Reuter:2005kb}, while in other studies the RG scale was linked
to the fourth root of the energy density $\rho$ \cite{Guberina:2002wt}, the cosmological
event/particle horizons \cite{Bauer:2005rpa}, or curvature invariants like Ricci scalar
\cite{Frolov:2011ys}, \cite{Copeland:2013vva}. Similarly, a law for $G(k)$ together with a scaling
relation will provide the evolution of $\rho$ from (\ref{kiyetn}).

We will consider that the cosmology at early times corresponds to very high energies, where the
energy-dependent couplings are given by the well-known ultraviolet (UV) non-Gaussian fixed point
of the RG flow evolution
\begin{equation}
\Lambda=\lambda_{\ast}k^{2}\,\,\,\,\,\,,\,\,\,\,\,\, G=\frac{g_{\ast}}{k^{2}}\,,
\label{scaling}
\end{equation}
where $\lambda_{\ast},g_{\ast}>0$ are dimensionless constants (note that the above scalings are
also consistent with dimensional analysis without the introduction of a new energy scale).
The vacuum solution in the early times has been studied in \cite{2016PhRvD..94j3514K}.
Here, we will study the cosmological evolution when the universe is in the radiation era
described by a perfect fluid with the relativistic equation of state parameter $w=1/3$. Using this
equation of state and the UV law (\ref{scaling}), equation (\ref{kiyetn}) or (\ref{kiyet}) gives
\begin{equation}
\rho=\frac{c\bar{\Lambda}}{g_{\ast}\lambda_{\ast}a^{4}}\,.
\label{djhhh}
\end{equation}
This is the same evolution for the energy density as in the standard FRW cosmology, what is
interesting because this equation is a basic element for the successful description of the early
thermal history of the universe. So, although in the non-conservation equation (\ref{hyeh}) there
are correction terms, in the high energy radiation regime, these terms cancel each other and the
simple dilution law (\ref{djhhh}) arises.

\subsection{Scaling $k\propto 1/t$}
\label{inverse time}

In cosmological models of Asymptotically Safe gravity it is common to use as a reasonable scaling
the following expression \cite{Bonanno:2001xi}
\begin{equation}
k=\frac{\xi}{t}\,,
\label{scale1a}
\end{equation}
where $\xi>0$ is a dimensionless parameter and time $t$ is considered positive valued. When $k$
is in the high energy regime, the time $t$ takes sufficiently small values, where we are
interested to understand the cosmological behaviour. From (\ref{scale1a}) obviously $k$ decreases
with time.

The relation of $\psi$ with time $t$ is $e^{\psi}=\lambda_{\ast}\xi^{2}/(\bar{\Lambda}t^{2})$ and
the Friedmann equation (\ref{hywte}) becomes
\begin{equation}
H^{2}+\frac{\kappa}{a^{2}}=\Big(\frac{\lambda_{\ast}\xi^{2}}{3}\!-\!1\Big)\frac{1}{t^{2}}
+\frac{2H}{t}+\frac{8\pi c\bar{\Lambda}t^{2}}{3\lambda_{\ast}\xi^{2}a^{4}}\,.
\label{drlvy}
\end{equation}
Due to the explicit time dependence in equation (\ref{drlvy}) and the positiveness of (\ref{scale1a}),
the generated cosmic evolution is not time-reversible, as it is also seen from the form and the
analysis of the solutions below. Equation (\ref{drlvy}) is invariant under the transformation
$t\rightarrow \lambda t$, $a\rightarrow \lambda a$, therefore defining
\begin{equation}
u=\frac{a}{t}\,,
\label{nergh}
\end{equation}
we find the equation
\begin{equation}
\dot{u}^{2}=\frac{1}{t^{2}}\Big(\omega^{2}u^{2}+\frac{\sigma^{2}}{u^{2}}-\kappa\Big)\,,
\label{ynsn}
\end{equation}
where
\begin{equation}
\omega=\sqrt{\frac{\lambda_{\ast}\xi^{2}}{3}}\,\,\,\,\,,\,\,\,\,\,
\sigma=\sqrt{\frac{8\pi c\bar{\Lambda}}{3\lambda_{\ast}\xi^{2}}}\,\,.
\label{gywg}
\end{equation}
So, while $\omega$ is a dimensionless combination of parameters of the theory, the constant
$\sigma$ carries dimensions of inverse length squared and is basically a rephrase of the integration
constant $c$. In terms of $\omega,\sigma$, the evolution equation (\ref{djhhh}) for $\rho$ takes
the form
\begin{equation}
\rho=\frac{9\omega^{2}\sigma^{2}}{8\pi g_{\ast}\lambda_{\ast}a^{4}}\,.
\label{djbhu}
\end{equation}

A direct integration of (\ref{ynsn}) gives the general solution
\begin{equation}
a(t)=\frac{\sqrt{\sigma}\,\,t}{\sqrt{2\omega}}\sqrt{\epsilon\Big(\frac{t}{t_{o}}\Big)^{\!\!\pm
2\omega}+\epsilon\Big(\frac{\kappa^{2}}{4\omega^{2}\sigma^{2}}\!-\!1\Big)
\Big(\frac{t_{o}}{t}\Big)^{\!\!\pm 2\omega}+\frac{\kappa}{\omega\sigma}}\,\,,
\label{kiefr}
\end{equation}
where $t_{o}>0$ is integration constant with dimensions of time. The upper sign of the $\pm$
(or $\mp$) symbol is characterized by the property $Ht>1$ (and gives only expanding solutions), while
the corresponding lower sign is characterized by $Ht<1$ (and contains also the case of contraction
with $H<0$). The sign $\epsilon=\text{sgn}\big[(\frac{t}{t_{o}})^{\pm 4\omega}\!-\!
\frac{\kappa^{2}}{4\omega^{2}\sigma^{2}}+\!1\big]$ is another sign symbol different
than the $\pm$ appeared in (\ref{kiefr}) and, depending on its value ($\epsilon=1$ or $\epsilon=-1$),
constraints appear on the domain of time $t$ of the corresponding branch. In addition, the quantity
inside the square root of (\ref{kiefr}) should be positive. The positivity of the right hand side
of (\ref{ynsn}) is trivially satisfied for the solution (\ref{kiefr}). In total, the solution
(\ref{kiefr}) contains two essential integration constants: one is $\sigma$ which is related to
matter and the second is $t_{o}$ which appears due to the explicit time dependence in equation
(\ref{drlvy}) of Hubble evolution (so, $t_{o}$ here is not the analogue of an absorbable
time-translation parameter). We can also obtain the evolution of the Hubble parameter
$H=\frac{1}{t}+\frac{\dot{u}}{u}=
\frac{1}{t}\Big(\!1\!\pm\!\sqrt{\omega^{2}\!-\!\frac{\kappa}{u^{2}}\!+\!\frac{\sigma^{2}}{u^{4}}}\Big)$
as
\begin{equation}
H(t)=\frac{(1\!\pm\!\omega)\big(\frac{t}{t_{o}}\big)^{\!\pm 2\omega}
+(1\!\mp\!\omega)\big(\frac{\kappa^{2}}{4\omega^{2}\sigma^{2}}\!-\!1\big)
\big(\frac{t_{o}}{t}\big)^{\!\pm 2\omega}+\frac{\epsilon\kappa}{\omega\sigma}}
{t\,\big[\big(\frac{t}{t_{o}}\big)^{\!\pm 2\omega}
+\big(\frac{\kappa^{2}}{4\omega^{2}\sigma^{2}}\!-\!1\big)
\big(\frac{t_{o}}{t}\big)^{\!\pm 2\omega}+\frac{\epsilon\kappa}{\omega\sigma}\big]}\label{uwkdod}\,,
\end{equation}
while the acceleration/deceleration of the universe is governed by the equation
$\frac{\ddot{a}}{a}=\frac{1}{t^{2}}\Big(\!\omega^{2}\!-\!\frac{\sigma^{2}}{u^{4}}\!\pm\!
\sqrt{\omega^{2}\!-\!\frac{\kappa}{u^{2}}\!+\!\frac{\sigma^{2}}{u^{4}}}\Big)$, found by
combining (\ref{hywte}), (\ref{hwynqo}). The 4-dimensional Ricci scalar is
$R=6\big(\frac{\ddot{a}}{a}+H^{2}+\frac{\kappa}{a^{2}}\big)=6\frac{1+2\omega^{2}}{t^{2}}\pm
6\frac{1+2t}{t^{2}}\sqrt{\omega^{2}\!-\!\frac{\kappa}{u^{2}}\!+\!\frac{\sigma^{2}}{u^{4}}}$.
In the case of spatially flat cosmology $\kappa=0$, the solution (\ref{kiefr}) is written as
\begin{equation}
a(t)=\frac{\sqrt{\sigma}\,\,t}{\sqrt{2\omega}}\,\sqrt{\Big(\frac{t}{t_{o}}\Big)^{\!\pm 2\omega}
\!-\!\Big(\frac{t_{o}}{t}\Big)^{\!\pm 2\omega}}\label{ufkaod}\,,
\end{equation}
where the upper $\pm$ branch holds for $t>t_{o}$, while the lower one for $t<t_{o}$.

\vspace{0.4cm}
$\uwave{\,\textit{Analysis of the solutions}\,}$ :
The behaviour of the solutions (\ref{kiefr}) depends on the value of the quantity $\omega\sigma$,
i.e. if $\omega\sigma$ is larger or smaller than $|\kappa|/2$.

$\bullet\,\,\,\,$ $\omega\sigma\geq \frac{|\kappa|}{2}\,.$
\newline
For both $\pm$ branches it is $\epsilon=1$. For the upper branch it arises that
$t\geq t_{1}=t_{o}\big(1\!-\!\frac{\kappa}{2\omega\sigma}\big)^{\!\frac{1}{2\omega}}$, while for the
lower branch it is
$t\leq t_{2}=t_{o}\big(1\!-\!\frac{\kappa}{2\omega\sigma}\big)^{\!-\frac{1}{2\omega}}$
(the spatially flat topology $\kappa=0$ is included in the present analysis with $t_{1}=t_{2}=t_{o}$
since $\omega\sigma>0$).
The {\textit{upper branch (first solution)}} starts expanding from $a=0$ at $t=t_{1}$ through a
decelerating era and passes into acceleration up to infinite scale factor, which at large $t$ it obeys
$a=\frac{t_{o}\sqrt{\sigma}}{\sqrt{2\omega}}(\frac{t}{t_{o}})^{1+\omega}\,,\,$
$H=\frac{1+\omega}{t}\,,\,$
$\frac{\ddot{a}}{a}=\frac{\omega(1+\omega)}{t^{2}}$; this acceleration could be examined if it
helps to the early inflationary era of the universe. The behaviour of the {\textit{lower branch
(second solution)}} is more complicated, depending on the value of $\omega$, and is described in the
following. If $\omega<1$, then the universe starts expanding from $a=0$ at $t=0$, where close to
$t=0$ it is $a=\frac{t_{o}\sqrt{\sigma}}{\sqrt{2\omega}}(\frac{t}{t_{o}})^{1-\omega}$,
$H=\frac{1-\,\omega}{t}$, while at some point during the evolution the expansion turns into
contraction which ends to $a=0$ at $t=t_{2}$; the universe is always decelerating with initial
deceleration close to $t=0$ of the form $\frac{\ddot{a}}{a}=-\frac{\omega(1-\,\omega)}{t^{2}}$.
If $\omega>1$, then the universe starts collapsing from infinite volume at $t=0$ until $a=0$ at
$t=t_{2}$, where close to $t=0$ it is
$a=\frac{t_{o}\sqrt{\sigma}}{\sqrt{2\omega}}(\frac{t}{t_{o}})^{\omega-1}$,
$H=-\frac{\omega-1}{t}$; initially, the evolution is accelerating, where close to $t=0$ it is
$\frac{\ddot{a}}{a}=\frac{\omega(\omega-1)}{t^{2}}$, while at some point passage into
deceleration occurs until $t=t_{2}$. If $\omega=1$ and $\kappa=-1,0$, then the universe starts
collapsing from a finite $a=\frac{t_{o}\sqrt{\sigma}}{\sqrt{2\omega}}$ at $t=0$ until $a=0$
at $t=t_{2}$ and it is always decelerating; for $\omega=1$ and $\kappa=1$, there
is an initial accelerating expansion from $a=\frac{t_{o}\sqrt{\sigma}}{\sqrt{2\omega}}$ at $t=0$,
which finally passes into a decelerating collapse until $a=0$ at $t=t_{2}$. The basic qualitative
features of the possible cosmic evolutions of the two branches with
$\omega\sigma\geq \frac{|\kappa|}{2}$, described above, are depicted in Fig. 1.

\begin{figure}[h!]
\centering
\begin{tabular}{cc}
\includegraphics*[width=160pt, height=130pt]{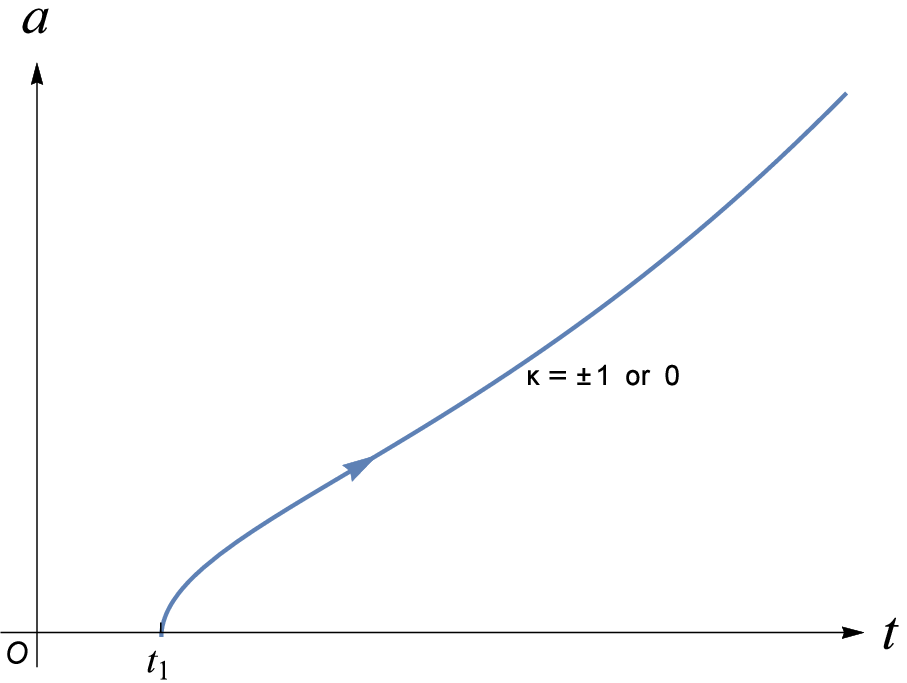}&
\hspace{2cm}
\includegraphics*[width=180pt, height=130pt]{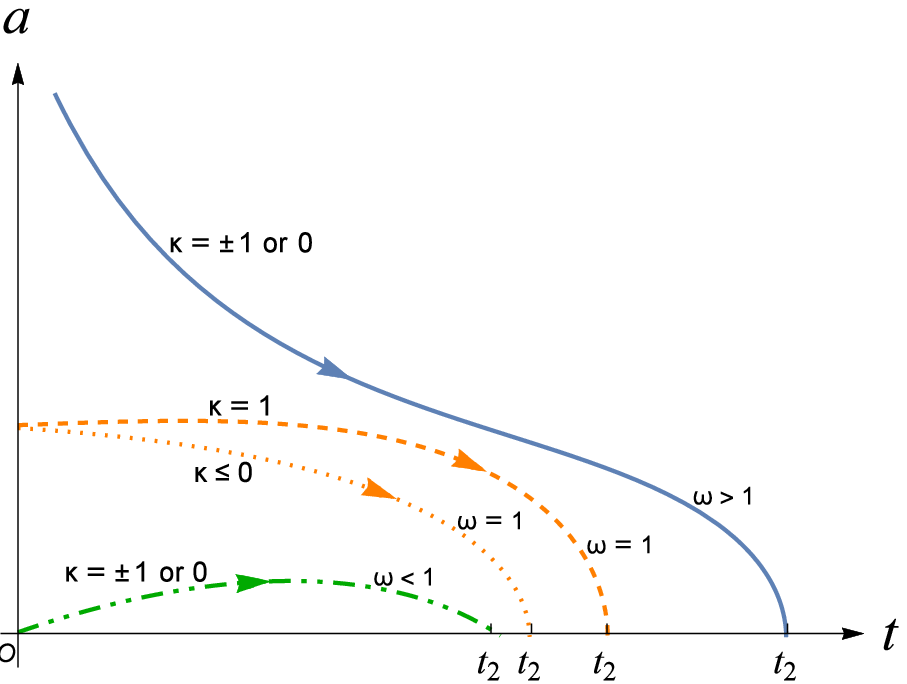}\\
\color{brown}{upper branch}   & \hspace{2cm} \color{brown}{lower branch}
\end{tabular}
\caption{Scale factor evolution for $\omega\sigma\geq \frac{|\kappa|}{2}$.}
\end{figure}

As an indication for the behaviour of the horizons, we will make for the above solutions
an analysis of the particle and event horizons. The particle horizon of a cosmological model is
defined by the comoving radius
$\mathlarger{\mathlarger{\chi}}=\int_{0}^{r}\!\frac{dr}{\sqrt{1-\kappa r^{2}}}$ of the observable
universe, $\mathlarger{\mathlarger{\chi}}_{_{\text{PH}}}(t)=\int_{t_{i}}^{t}\frac{dt}{a(t)}
=\int_{a_{i}}^{a}\!\frac{da}{Ha^{2}}$ (where $t_{i},a_{i}$ denote the initial time and initial scale
factor of the universe). The quantity $\mathlarger{\mathlarger{\chi}}_{_{\text{PH}}}$ is an increasing
function of $t$; of course, if the above integrals diverge close to $t_{i},a_{i}$, there is no
particle horizon. On the contrary, the event horizon is defined by
$\mathlarger{\mathlarger{\chi}}_{_{\text{EH}}}(t)=\int_{t}^{t_{\!f}}\frac{dt}{a(t)}=
\int_{a}^{a_{\!f}}\!\frac{da}{Ha^{2}}$ (where $t_{\!f},a_{\!f}$ denote the final time and final
scale factor of the universe)\footnote{In terms of the redshift $z$ it is
$\mathlarger{\mathlarger{\chi}}_{_{\text{PH}}}(z)=\frac{1}{a_{\ast}}\int_{z}^{z_{i}}\!
\frac{dz}{H(z)}$\,,
$\mathlarger{\mathlarger{\chi}}_{_{\text{EH}}}(z)=\frac{1}{a_{\ast}}\int_{z_{\!f}}^{z}\!
\frac{dz}{H(z)}$, where $z_{i},z_{\!f}$ is the initial, final redshift and $a_{\ast}$ is the today
scale factor.}. The quantity $\mathlarger{\mathlarger{\chi}}_{_{\text{EH}}}$ is a decreasing function
of $t$; again, if these integrals diverge close to $t_{\!f},a_{\!f}$, there is no event horizon. If
particle or event horizons exist, the corresponding proper distances of these horizons are
$d_{_{\text{P,PH}}}(t)=a(t)\mathlarger{\mathlarger{\chi}}_{_{\text{PH}}}(t)$,
$d_{_{\text{P,EH}}}(t)=a(t)\mathlarger{\mathlarger{\chi}}_{_{\text{EH}}}(t)$. Since our solutions
are early-times solutions, we are interested in the existence of particle horizons.
For the upper branch it arises from (\ref{kiefr}) that
$\mathlarger{\mathlarger{\chi}}_{_{\text{PH}}}(t)=\frac{1}{\sqrt{2\omega\sigma}}
\int_{_{\!(\!\frac{t_{i}}{t_{o}}\!)^{^{2\omega}}}}
^{(\!\frac{t}{t_{o}}\!)^{2\omega}}\!\frac{d\mathcal{T}}{\sqrt{\mathcal{T}
(\mathcal{T}+\frac{\kappa}{2\omega\sigma}-1)(\mathcal{T}+\frac{\kappa}{2\omega\sigma}+1)}}$,
where $\mathcal{T}=\big(\frac{t}{t_{o}}\big)^{\!2\omega}$.
This solution starts expanding from $a_{i}=0$ at $t_{i}=t_{1}$,
$\mathcal{T}_{i}=1-\frac{\kappa}{2\omega\sigma}$,
thus for $\mathcal{T}$ close to $\mathcal{T}_{i}$ the previous integral, defining
$\mathlarger{\mathlarger{\chi}}_{_{\text{PH}}}$, behaves as
$\sqrt{\mathcal{T}-\mathcal{T}_{i}}$ and is convergent, therefore particle horizon exists.
For the lower branch we consider the initially expanding solution with $\omega<1$, where
$\mathlarger{\mathlarger{\chi}}_{_{\text{PH}}}(t)=\frac{1}{\sqrt{2\omega\sigma}}
\int_{_{\!(\!\frac{t_{o}}{t}\!)^{^{2\omega}}}}
^{(\!\frac{t_{o}}{t_{i}}\!)^{2\omega}}\!\frac{d\textsf{T}}{\sqrt{\textsf{T}
(\textsf{T}+\frac{\kappa}{2\omega\sigma}-1)(\textsf{T}+\frac{\kappa}{2\omega\sigma}+1)}}$
with $\textsf{T}=\big(\frac{t_{o}}{t}\big)^{\!2\omega}$.
This solution starts expanding from $a_{i}=0$ at $t_{i}=0$, $\textsf{T}_{i}=\infty$,
thus for $\textsf{T}$ close to $\textsf{T}_{i}$ the previous integral, defining
$\mathlarger{\mathlarger{\chi}}_{_{\text{PH}}}$, behaves as $\textsf{T}^{-1/2}$
and is convergent, therefore a particle horizon also exists. This result, concerning
the $\omega<1$ solution of the lower branch, also arises directly from the behaviour
$a\sim t^{1-\omega}$ close to the origin. To compare with General Relativity, for a power-law
solution $a(t)\sim t^{\gamma}$ close to the big bang $a_{i}=0$, $t_{i}=0$, there is particle horizon
for $\gamma<1$, while there is no particle horizon for $\gamma\geq 1$.

$\bullet\,\,\,\,$ $\omega\sigma<\frac{|\kappa|}{2}\,.$
\newline
Here the analysis is different for the two possible cases $\kappa=-1$ and $\kappa=1$ of spatial
topology.

Starting with $\dashuline{\,\textit{$\kappa=-1$}\,}$, where the situation is simpler since there is
only one solution for each $\pm$ branch, it arises that $\epsilon=1$. These solutions are
reminiscent to the solutions above with $\omega\sigma\geq \frac{|\kappa|}{2}\,.$ So, for the upper
branch it arises that
$t\geq t_{1}=t_{o}\big(1\!-\!\frac{\kappa}{2\omega\sigma}\big)^{\!\frac{1}{2\omega}}$, while for the
lower branch it is
$t\leq t_{2}=t_{o}\big(1\!-\!\frac{\kappa}{2\omega\sigma}\big)^{\!-\frac{1}{2\omega}}$.
The {\textit{upper branch (first solution)}} starts expanding from $a=0$ at $t=t_{1}$ through a
decelerating era and passes into acceleration up to infinite scale factor, which at large $t$ it obeys
$a=\frac{t_{o}\sqrt{\sigma}}{\sqrt{2\omega}}(\frac{t}{t_{o}})^{1+\omega}\,,\,$
$H=\frac{1+\omega}{t}\,,\,$
$\frac{\ddot{a}}{a}=\frac{\omega(1+\omega)}{t^{2}}$; this acceleration could be examined if it
helps to the early inflationary era of the universe. The behaviour of the {\textit{lower branch
(second solution)}} is more
complicated, depending on the value of $\omega$, and is described in the following. If $\omega<1$,
then the universe starts expanding from $a=0$ at $t=0$, where close to $t=0$ it is
$a=\frac{t_{o}\sqrt{\sigma}}{\sqrt{2\omega}}(\frac{t}{t_{o}})^{1-\omega}$, $H=\frac{1-\,\omega}{t}$,
while at some point during the evolution the expansion turns into contraction which ends to $a=0$
at $t=t_{2}$; the universe is always decelerating with initial deceleration close to $t=0$ of the
form $\frac{\ddot{a}}{a}=-\frac{\omega(1-\,\omega)}{t^{2}}$. If $\omega>1$, then the universe starts
collapsing from infinite volume at $t=0$ until $a=0$ at $t=t_{2}$, where close to $t=0$ it is
$a=\frac{t_{o}\sqrt{\sigma}}{\sqrt{2\omega}}(\frac{t}{t_{o}})^{\omega-1}$,
$H=-\frac{\omega-1}{t}$; initially, the evolution is accelerating, where close to $t=0$ it is
$\frac{\ddot{a}}{a}=\frac{\omega(\omega-1)}{t^{2}}$, while at some point passage into
deceleration occurs until $t=t_{2}$. If $\omega=1$, then the universe starts collapsing at $t=0$
from a finite $a=\frac{t_{o}\sqrt{\sigma}}{\sqrt{2\omega}}$ with $H=0$ until $a=0$ at $t=t_{2}$
and it is decelerating at all times. The basic qualitative features of the possible cosmic
evolutions of the two branches with $\omega\sigma<\frac{|\kappa|}{2}$ and $\kappa=-1$, described
above, are depicted in Fig. 2.

\begin{figure}[h!]
\centering
\begin{tabular}{cc}
\includegraphics*[width=160pt, height=130pt]{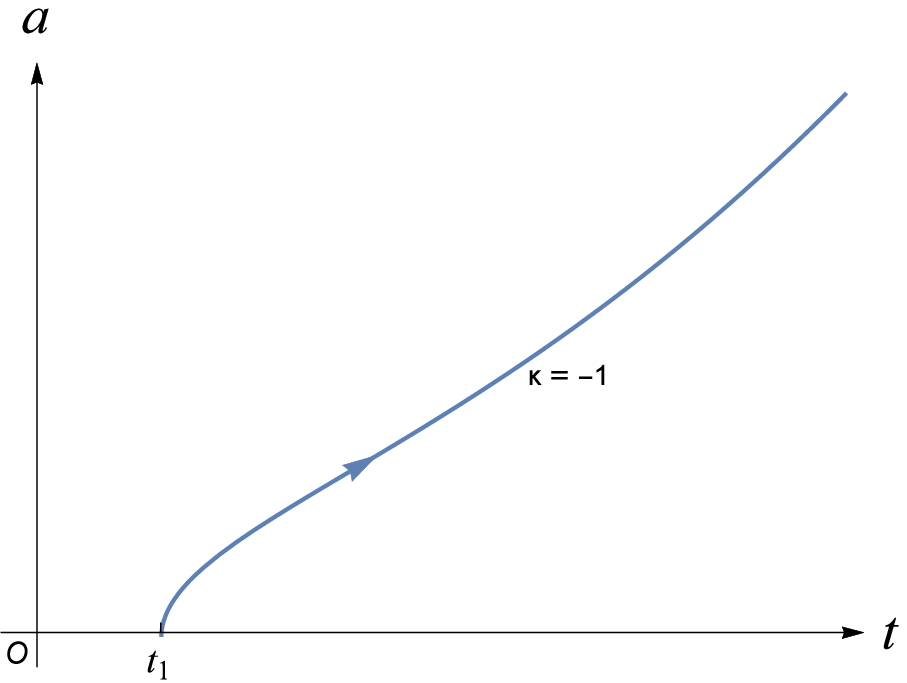}&
\hspace{2cm}
\includegraphics*[width=180pt, height=130pt]{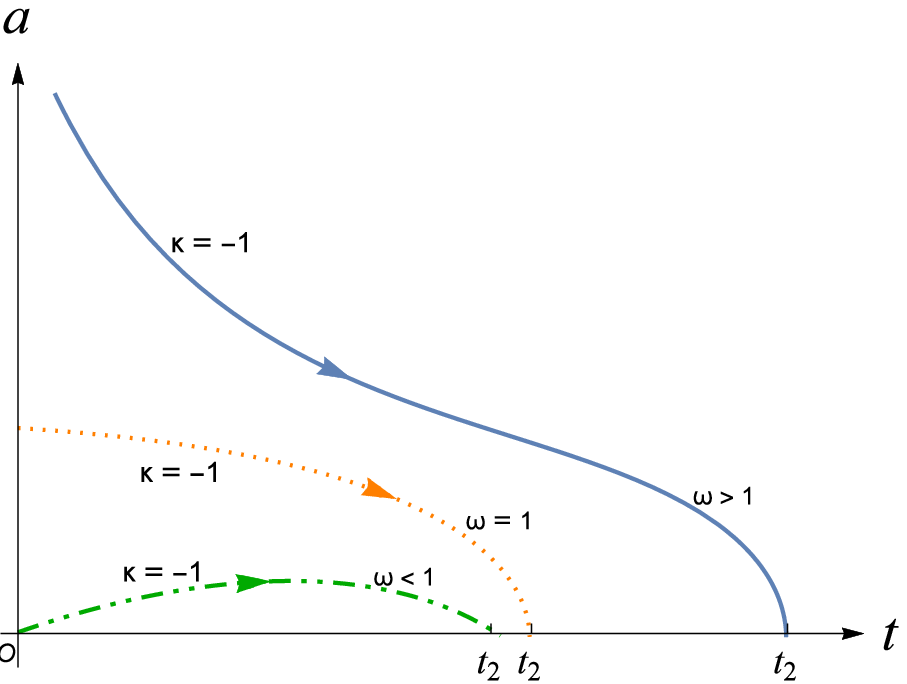}\\
\color{brown}{upper branch}   & \hspace{2cm} \color{brown}{lower branch}
\end{tabular}
\caption{Scale factor evolution for $\omega\sigma< \frac{|\kappa|}{2}$, $\kappa=-1$.}
\end{figure}

We now pass to the analysis of the universe evolution with $\dashuline{\,\,\textit{$\kappa=1$}\,\,}$.
The two $\pm$ branches appeared throughout are due to the presence of $\dot{u}$ squared in equation
(\ref{ynsn}). Depending on the form of the various branches-solutions, there are two alternatives:
first, these branches stand as they are and they define maximally extended solutions, where no
physical matching between different branches can occur; or second, the appropriate branches should
match together so as to define the maximally extended physically meaningful solution. All the above
solutions described up to now belong to the first alternative, i.e. they make sense as independent
branches and nothing more can be said. The reason is that at the end points $t=t_{1}$ or $t=t_{2}$
of the various branches, the first derivative $\dot{a}$ diverges to $+\infty$ or $-\infty$
respectively, therefore no notion of smooth passage can be defined. However, the solutions of the
present case with $\kappa=1$, to be described below, belong to the second alternative and the total
physical solutions arise after the matching of the appropriate branches.
\newline
We start with the upper $\pm$ branch, where there are now two different solutions. The first one has
$t\geq t_{3}=t_{o}\big(\frac{\kappa^{2}}{4\omega^{2}\sigma^{2}}\!-\!1\big)^{\!\frac{1}{4\omega}}$
and $\epsilon=1$. This solution describes a universe which starts expanding at $t=t_{3}$ from a
finite scale factor with $H=\frac{1}{t_{3}}$ and a finite Ricci scalar $R$; the expansion is always
accelerating and continues to infinite volume, where at large $t$ it obeys
$a=\frac{t_{o}\sqrt{\sigma}}{\sqrt{2\omega}}(\frac{t}{t_{o}})^{1+\omega}\,,\,$
$H=\frac{1+\omega}{t}\,,\,$ $\frac{\ddot{a}}{a}=\frac{\omega(1+\omega)}{t^{2}}$;
this acceleration could be examined if it helps to the early inflationary era of the universe.
The second solution occurs for $t_{1}'\leq t\leq t_{3}$, where
$t_{1}'=t_{o}\big(\frac{\kappa}{2\omega\sigma}\!-\!1\big)^{\!\frac{1}{2\omega}}$,
and it has $\epsilon=-1$. It starts with an accelerated expansion from $a=0$ at $t=t_{1}'$ and
passes into deceleration until the end point $t=t_{3}$, where the scale factor is finite and
$H=\frac{1}{t_{3}}$. This solution has the extra interesting feature that the acceleration is
transient, so if this acceleration is managed to be interpreted as early-times inflation, there
is a natural exit from this.
\newline
We continue with the analysis of the lower $\pm$ branch, where still there are two
different solutions, but their behaviours are more complicated because they depend on the value
of $\omega$, and are described in the following. The first of these solutions has $t\leq t_{4}=t_{o}
\big(\frac{\kappa^{2}}{4\omega^{2}\sigma^{2}}\!-\!1\big)^{\!-\frac{1}{4\omega}}$ and
$\epsilon=1$. For $\omega<1$, it starts expanding from $a=0$ at $t=0$, where close to $t=0$ it is
$a=\frac{t_{o}\sqrt{\sigma}}{\sqrt{2\omega}}(\frac{t}{t_{o}})^{1-\omega}$,
$H=\frac{1-\omega}{t}$, $\frac{\ddot{a}}{a}=-\frac{\omega(1-\omega)}{t^{2}}$; the evolution
starts with a deceleration era and passes into acceleration until the end point $t=t_{4}$, where
the scale factor is finite and $H=\frac{1}{t_{4}}$.
For $\omega>1$ the solution is bouncing; the universe starts collapsing from
infinite volume at $t=0$ up to some time, and then, it turns into expansion until the end point
$t=t_{4}$, where the scale factor is finite and $H=\frac{1}{t_{4}}$; close to $t=0$ it is
$a=\frac{t_{o}\sqrt{\sigma}}{\sqrt{2\omega}}(\frac{t_{o}}{t})^{\omega-1}$,
$H=-\frac{\omega-1}{t}$, $\frac{\ddot{a}}{a}=\frac{\omega(\omega-1)}{t^{2}}$, while
the evolution is accelerating at all times. If $\omega=1$, the solution is expanding and
accelerating, it starts at $t=0$ from a finite $a=\frac{t_{o}\sqrt{\sigma}}{\sqrt{2\omega}}$
with $H=0$ and ends at $t=t_{4}$ at a finite scale factor with $H=\frac{1}{t_{4}}$.
The second solution occurs for $t_{4}\leq t\leq t_{2}'$, where
$t_{2}'=t_{o}\big(\frac{\kappa}{2\omega\sigma}\!-\!1\big)^{\!-\frac{1}{2\omega}}$, and it has
$\epsilon=-1$. This solution starts expanding at $t=t_{4}$ with a finite scale factor,
$H=\frac{1}{t_{4}}$ and a finite Ricci scalar; after some time the expansion turns into
contraction until the end point $a=0$ at $t=t_{2}'$. For $\omega\leq 1$ there
is deceleration at all times; for $\omega>1$ there is an initial deceleration which passes into
acceleration.
\newline
After having described the various branches for $\kappa=1$, we now come to the appropriate matchings
in order to define the possible total solutions. Each branch has an integration constant $t_{o}$.
However, the differential equation (\ref{ynsn}) is only of first order, so only a single
integration constant $t_{o}$ should describe a total solution of equation (\ref{ynsn})
instead of the two different $t_{o}$'s of the two branches which are matched together.
Therefore, the following two total solutions are the only ones which arise, the first with
$\epsilon=1$ and the second with $\epsilon=-1$.
\newline
\textit{Total solution 1.} The first solution described above (which we call here with respect
to the time arrow as left) of the lower branch matches naturally and smoothly with the
first solution (named as right) of the upper branch. Namely, if $t_{o-}$ and $t_{o+}$
denote the two corresponding integration constants of the left and right solutions
respectively and it holds $t_{o+}=t_{o-}
\big(\frac{\kappa^{2}}{4\omega^{2}\sigma^{2}}\!-\!1\big)^{\!-\frac{1}{2\omega}}$, then the two
solutions match at $t_{4}=t_{3}$. At the matching point, it can be seen{\footnote{At the matching
point, one sets $t_{3}=t_{4}$ and the solution (\ref{kiefr}) implies that $a$ is continuous. Then,
the expression for $H$ given just before equation (\ref{uwkdod}), or equation (\ref{uwkdod})
itself, provides that $H$ is also continuous with value $t_{3}^{-1}=t_{4}^{-1}$ at the matching
point (this value has been mentioned above, in the analysis of the branches). These results are
sufficient to guarantee the full smoothness of $a(t)$ since progressive higher differentiations
of (\ref{drlvy}) generate all the higher derivatives of $H$ only at linear order; therefore, these
derivatives can uniquely be solved algebraically at the matching point in terms of lower derivatives,
and as a result, these higher derivatives are also continuous. Since all the higher derivatives
$a^{(n)}$ of $a$ are expressed in terms of higher derivatives of $H$, thus $a^{(n)}$ are also
continuous at the matching point.}}
that not only the scale factor $a$, but also $H$, $\frac{\ddot{a}}{a}$ and all the higher derivatives
of $a(t)$ coincide among the two sides, therefore there is an infinitely smooth passage from one
branch to the other and the solution is infinitely smooth. The total solution, which consists of
the two separate branches, is a solution that starts at $t=0$ and continues up to infinite time,
possessing the characteristics of each branch in its domain of validity. So, for $\omega<1$ there
is a continuous expansion from $a=0$ up to infinite volume, and the solution is initially decelerating
and finally accelerating; for $\omega>1$ there is a bouncing solution coming from infinite volume up
to a minimum scale factor, from where it expands again to infinity, and the solution is always
accelerating; for $\omega=1$ the universe starts expanding from a finite scale factor up to infinite
volume and the solution is always accelerating.
\newline
\textit{Total solution 2.} The second solution (left one) of the upper branch and the second
solution (right one) of the lower branch also match together. Now, the integration constant of the
right branch should be $t_{o+}=t_{o-}
\big(\frac{\kappa^{2}}{4\omega^{2}\sigma^{2}}\!-\!1\big)^{\!\frac{1}{2\omega}}$ and the matching
occurs at $t_{3}=t_{4}$. Also here, at the matching point, the solution is infinitely smooth, so
all derivatives of $a(t)$ coincide for the two branches. For the total cosmic evolution, the
scale factor starts from $a=0$ at $t_{1}'$ with acceleration, it reaches a maximum and then
recollapses to $a=0$ at $t_{2}'$ either with deceleration (for $\omega\leq 1$) or with acceleration
(for $\omega> 1$). Irrespectively of the recollapsing behaviour of the total solution, the initial
temporary accelerating phase may be of some special interest and attention concerning the inflation
and exit from inflation question in the early universe.
\newline
The basic qualitative features of the possible cosmic evolutions of the two total solutions with
$\omega\sigma<\frac{|\kappa|}{2}$ and $\kappa=1$, described above, are depicted in Fig. 3.

\begin{figure}[h!]
\centering
\begin{tabular}{cc}
\includegraphics*[width=180pt, height=130pt]{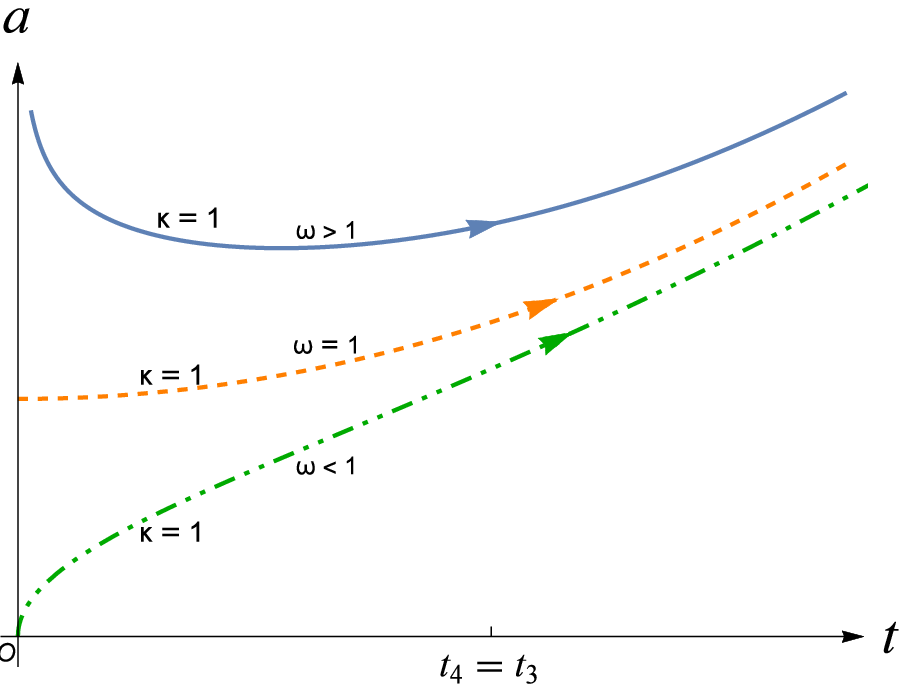}&
\hspace{2cm}
\includegraphics*[width=180pt, height=130pt]{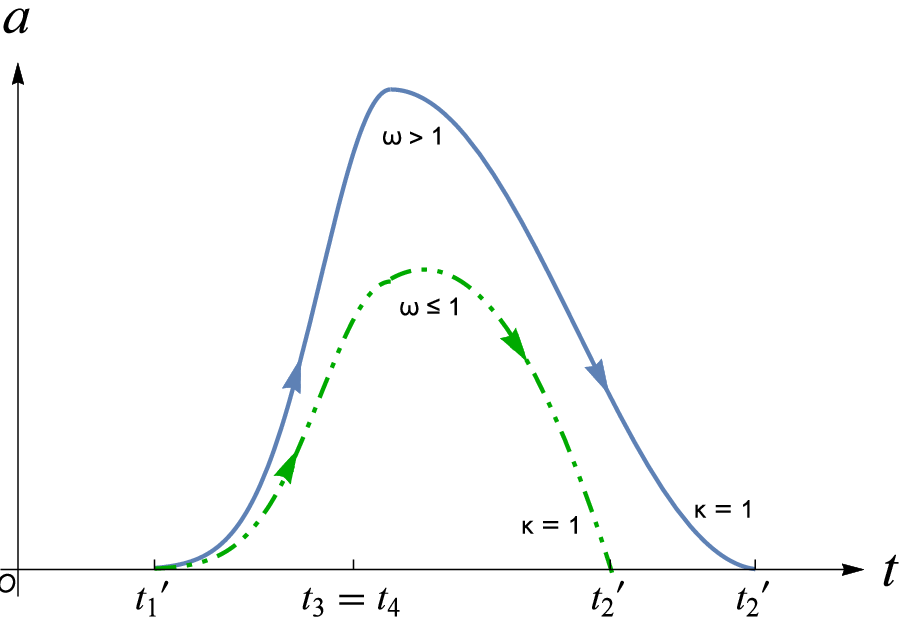}\\
\color{brown}{total solution 1}   & \hspace{2cm} \color{brown}{total solution 2}
\end{tabular}
\caption{Scale factor evolution for $\omega\sigma< \frac{|\kappa|}{2}$, $\kappa=1$.}
\end{figure}

Note that all solutions obtained in this subsection with $k\propto 1/t$ have an infinite
Ricci scalar $R$ whenever the scale factor $a$ vanishes. In this sense, the solutions which
start at $a=0$ with a big bang or result in $a=0$ with a big crunch can be considered as singular
with respect to the divergences of the curvature invariants.

To recap very briefly in order to encode the behaviour of all the solutions found above for the
scaling $k\propto 1/t$ : (i) there exist (for all spatial curvatures) strictly expanding or collapsing
solutions between zero and infinite scale factor, (ii) there are bouncing solutions (for positive
spatial curvature) from infinite to infinite scale factor, and finally, (iii) there are recollapsing
solutions (for all spatial curvatures) which start from a big bang and end up to a big crunch
(actually one of these, with positive spatial curvature, possesses an initial transient
accelerating era).

\subsection{Scaling $k\propto 1/a$}
\label{inverse scale factor}

An interesting scaling for all spacetimes in Asymptotic Safety would be of the form
$k=\frac{1}{d_{P}}$, where $d_{P}$ is some proper distance appropriately defined in each case.
For the cosmological metric discussed here, the proper distance refers to that of a point at
coordinate distance $r$ from the origin $r=0$ of the coordinate system (\ref{jkwk}). This is the
physical distance in spacetime given by the expression $d_{P}=a\chi$, where $\chi\geq 0$ is the
corresponding proper distance of the spatial part of the metric. The relation of $\chi$ and $r$ is
given by the known expression $r^{2}=\frac{1}{\kappa}\sin^{2}(\sqrt{\kappa}\,\chi)$, where for
$\kappa=1$ it is $\chi\leq \pi$. Therefore $k$ becomes $k=\frac{1}{a\chi}$. It is not quite
obvious in cosmology what should be the relevant value of the comoving distance $\chi$ entering this
equation. However, even for $\kappa\leq 0$, where $\chi$ extends to infinity, the relevant $\chi$
should be at most the one corresponding to the horizon distance. Therefore, we set
\begin{equation}
k=\frac{k_{0}}{a}\,,
\label{jdgjy}
\end{equation}
which is the scaling law to be studied, where $k_{0}>0$ is a fixed energy scale.
Since $k$ is in the high energy regime, the scale factor $a$ takes sufficiently
small values, where we are interested to understand the cosmological behaviour. The relation of
$\psi$ with scale factor $a$ is $e^{\psi}=\lambda_{\ast}k_{0}^{2}/(\bar{\Lambda}a^{2})$ and the
Friedmann equation (\ref{hywte}) turns out to be identically satisfied for any $a(t)$ given that
$\kappa=1=\frac{\lambda_{\ast}k_{0}^{2}}{3}+\frac{8\pi c\bar{\Lambda}}{3\lambda_{\ast}k_{0}^{2}}$.
Eventually, the above scaling $k\propto a^{-1}$ does not provide some useful information for the
early-times cosmology. The same outcome of phenomenologically unacceptable solutions for the
scaling (\ref{jdgjy}) also arises from other AS cosmology studies \cite{Bonanno:2001xi},
\cite{Reuter:2005kb}, \cite{Bonanno:2001hi}.

\subsection{Scaling $k\propto H$}
\label{hubble scale}

Another popular cutoff identification used in the literature is the Hubble scale, so that
$k\sim H(t)$  \cite{Reuter:2005kb}, namely
\begin{equation}
k=\xi H(t)\,,
\label{scale1}
\end{equation}
with $\xi$ a constant. In general, $H$ can be positive or negative during the cosmic evolution.
However, according to the interpretation of $k$ as the inverse of the characteristic scale over
which the averaging procedure is performed, we assume the positivity of $k$. For this reason, the
dimensionless parameter $\xi$ is $\xi>0$ for $H>0$ and $\xi<0$ for $H<0$. Since $\xi$ is a fixed
parameter of the theory, it cannot in principle change sign and flip between a negative and a
positive value, so recollapsing or bouncing solutions are not possible here and a solution is either
always expanding or always contracting. As a result, the scale factor $a$ can play the role of a
``time'' parameter along all the cosmic evolution, not only for a segment of the evolution, and it
can be used in order to capture the qualitative features of a solution.

From (\ref{scale1}) it arises $e^{\psi}=\lambda_{\ast}\xi^{2}H^{2}/\bar{\Lambda}$ and the Friedmann
equation (\ref{hywte}) becomes
\begin{equation}
(1\!-\!\omega^{2})H^{2}+2\dot{H}+\frac{\dot{H}^{2}}{H^{2}}-\frac{\sigma^{2}}{a^{4}H^{2}}
+\frac{\kappa}{a^{2}}=0\,,
\label{hyd}
\end{equation}
where
\begin{equation}
\omega=\sqrt{\frac{\lambda_{\ast}\xi^{2}}{3}}\,\,\,\,\,,\,\,\,\,\,
\sigma=\sqrt{\frac{8\pi c\bar{\Lambda}}{3\lambda_{\ast}\xi^{2}}}\,\,.
\label{gyhg}
\end{equation}
So, as in the previous subsection with scaling $k\propto 1/t$, also here, while $\omega$
is a dimensionless combination of parameters of the theory, the constant $\sigma$ on the contrary
carries dimensions of inverse length squared and is basically a rephrase of the integration constant
$c$. In terms of $\omega,\sigma$, the evolution equation (\ref{djhhh}) for $\rho$ takes the form
\begin{equation}
\rho=\frac{9\omega^{2}\sigma^{2}}{8\pi g_{\ast}\lambda_{\ast}a^{4}}\,.
\label{djbng}
\end{equation}
Note that the Friedmann equation (\ref{hyd}) is of higher order (it contains $\dot{H}$) due to
the scaling law (\ref{scale1}), while the Raychaudhuri equation (\ref{hwynqo}) contains $\ddot{H}$.
Therefore, the first task is to integrate (\ref{hyd}) and find $H$. This is what we are going to do
in the following finding $H$ as a function of $a$. The next step would be the integration of this
expression $H(a)$ in order to obtain $a(t)$, although this is not necessary in order to understand
the basic characteristics of the cosmic evolution.

Equation (\ref{hyd}) is written as
\begin{equation}
a^{2}\ddot{a}^{2}-\omega^{2}\dot{a}^{4}+\kappa\dot{a}^{2}-\sigma^{2}=0\,.
\label{heu}
\end{equation}
Setting
\begin{equation}
u=\dot{a}^{2}\,,
\label{deghu}
\end{equation}
equation (\ref{heu}) takes the form
\begin{equation}
\Big(\frac{du}{da}\Big)^{\!2}=\frac{4}{a^{2}}\big(\omega^{2}u^{2}\!-\!\kappa u+\sigma^2\big)\,.
\label{ikijm}
\end{equation}
Direct integration of (\ref{ikijm}) gives the general solution
\begin{equation}
u(a)=\frac{\sigma}{2\omega}\Big[\epsilon\Big(\frac{a}{a_{o}}\Big)^{\!\!\pm 2\omega}
+\epsilon\Big(\frac{\kappa^{2}}{4\omega^{2}\sigma^{2}}\!-\!1\Big)
\Big(\frac{a_{o}}{a}\Big)^{\!\!\pm 2\omega}+\frac{\kappa}{\omega\sigma}\Big]\,,
\label{jwkce}
\end{equation}
where $a_{o}>0$ is a dimensionless integration constant.

The Hubble parameter turns out from (\ref{deghu}), (\ref{jwkce}) to be
\begin{equation}
|H(a)|=\frac{\sqrt{\sigma}}{\sqrt{2\omega}\,a}
\,\sqrt{\epsilon\Big(\frac{a}{a_{o}}\Big)^{\!\!\pm 2\omega}
+\epsilon\Big(\frac{\kappa^{2}}{4\omega^{2}\sigma^{2}}\!-\!1\Big)
\Big(\frac{a_{o}}{a}\Big)^{\!\!\pm 2\omega}+\frac{\kappa}{\omega\sigma}}\,.
\label{hywodh}
\end{equation}
Since $\frac{du}{da}=2\ddot{a}$, the upper sign of the $\pm$ (or $\mp$) symbol is characterized
by acceleration ($\ddot{a}>0$), while the corresponding lower sign is characterized by deceleration
($\ddot{a}<0$). The sign
$\epsilon=\text{sgn}\big[(\frac{a}{a_{o}})^{\pm 4\omega}\!-\!
\frac{\kappa^{2}}{4\omega^2\sigma^2}\!+\!1\big]$ is another sign symbol different than the $\pm$
appeared in (\ref{jwkce}), (\ref{hywodh}) and, depending on its value ($\epsilon=1$ or $\epsilon=-1$),
constraints appear on the domain of the scale factor $a$ of the corresponding branch. In addition,
the quantity inside the square root of (\ref{hywodh}) should be positive. The positivity of the right
hand side of (\ref{ikijm}) is trivially satisfied for the solution (\ref{jwkce}). In total, the
solution (\ref{jwkce}) or (\ref{hywodh}) contains two essential integration constants: one is
$\sigma$ which is related to matter and the second is $a_{o}$ which appears due to the higher order
of Friedmann equation (\ref{hyd}). Integration of (\ref{hywodh}) can provide explicitly the
dependency of the scale factor with time, i.e. $a(t)$ and then also $H(t)$. Since the time $t$
appears in (\ref{hywodh}) only through $H$, or also in (\ref{hyd}) only through $H$ and its derivatives,
and not explicitly, a time-translation integration constant will always appear in a solution $a(t)$.
Moreover, equation (\ref{hyd}) is time-reversible, which is equivalent to the presence of the absolute
value in (\ref{hywodh}), showing that the collapsing solutions with $H(a)<0$ arise from the expanding
ones with $H(a)>0$ by just inverting the cosmic evolution backwards. However, a solution cannot
contain both expanding and contracting eras, thus possible expanding and collapsing branches cannot
be matched, and therefore we will only focus on the expanding solutions of (\ref{hywodh}), while the
corresponding contracting solutions can easily be realized by inversion. Among the expanding branches
of solutions, it may be possible some matching to exist in order to obtain the total cosmic evolution.
The acceleration/decelaration of the universe is governed by the equation
$\frac{\ddot{a}}{a}=\pm\frac{\sigma}{2a_{o}^{2}}\big|\big(\frac{a_{o}}{a}\big)^{\!2(1\mp\omega)}
-\big(\frac{\kappa^{2}}{4\omega^{2}\sigma^{2}}\!-\!1\big)
\big(\frac{a_{o}}{a}\big)^{\!2(1\pm\omega)}\big|$, while the 4-dimensional Ricci scalar is
$R\!=\!6\big(\frac{\ddot{a}}{a}+H^{2}+\frac{\kappa}{a^{2}}\big)
\!=\!\frac{3\epsilon\sigma}{\omega a_{o}^{2}}
\Big[(1\!\pm\!\omega)\big(\frac{a_{o}}{a}\big)^{\!2(1\mp\omega)}\!+\!
(1\!\mp\!\omega)\big(\frac{\kappa^{2}}{4\omega^{2}\sigma^{2}}\!-\!1\big)
\big(\frac{a_{o}}{a}\big)^{\!2(1\pm\omega)}
\!+\!\frac{\epsilon\kappa(1+2\omega^{2})}{\omega\sigma}\big(\frac{a_{o}}{a}\big)^{\!2}\Big]$.
In the case of spatially flat cosmology $\kappa=0$, the solution (\ref{hywodh}) is written as
\begin{equation}
|H(a)|=\frac{\sqrt{\sigma}}{\sqrt{2\omega}\,a}\sqrt{\Big(\frac{a}{a_{o}}\Big)^{\!\!\pm 2\omega}
\!-\!\Big(\frac{a_{o}}{a}\Big)^{\!\!\pm 2\omega}}\,,
\label{hywrdh}
\end{equation}
where the upper $\pm$ branch holds for $a>a_{o}$ (and is accelerating), while the lower one for
$a<a_{o}$ (and is decelerating).

\vspace{0.4cm}
$\uwave{\,\textit{Analysis of the solutions}\,}$ :
The behaviour of the solutions (\ref{hywodh}) depends on the value of the quantity $\omega\sigma$,
i.e. if $\omega\sigma$ is larger or smaller than $|\kappa|/2$.

$\bullet\,\,\,\,$ $\omega\sigma\geq \frac{|\kappa|}{2}\,.$
\newline
For both $\pm$ branches it is $\epsilon=1$. For the upper branch it arises that
$a\geq a_{1}=a_{o}\big(1\!-\!\frac{\kappa}{2\omega\sigma}\big)^{\!\frac{1}{2\omega}}$, while for the
lower branch it is
$a\leq a_{2}=a_{o}\big(1\!-\!\frac{\kappa}{2\omega\sigma}\big)^{\!-\frac{1}{2\omega}}$
(the spatially flat topology $\kappa=0$ is included in the present analysis with $a_{1}=a_{2}=a_{o}$
since $\omega\sigma>0$). The {\textit{upper branch (first solution)}} starts expanding from $a=a_{1}$
with $H=0$, $\dot{a}=0$, $\frac{\ddot{a}}{a}=\frac{\sigma}{a_{1}^{2}}$, up to infinite volume,
being always accelerating; this acceleration could be examined if it helps to the early inflationary
era of the universe. Since there is time translation symmetry, the time at the starting point of
the expansion can be considered as $t=0$. As it will be explained below, this branch cannot be
extended in the past and being matched to the lower branch. It can be seen that the Ricci scalar
$R$ at the starting point $a=a_{1}$ is finite; moreover, from equation (\ref{djbng}) the energy
density $\rho$ at $a_{1}$ is also finite, therefore this solution is clearly non-singular. The
{\textit{lower branch (second solution)}} starts expanding from $a=0$, where $H$, $\dot{a}$ diverge
to $+\infty$, while $\frac{\ddot{a}}{a}$ diverges to $-\infty$. The initial time can be set to be
$t=0$, and close to $t=0$ it is
$H=\frac{\sqrt{\sigma}}{\sqrt{2\omega}\,a_{o}}(\frac{a_{o}}{a})^{\omega+1}$. The expansion continues
until the end point $a=a_{2}$, where $H=0$, $\dot{a}=0$,
$\frac{\ddot{a}}{a}=-\frac{\sigma}{a_{2}^{2}}$, and the evolution is always decelerating.
The basic qualitative features of the possible cosmic evolutions of the two branches with
$\omega\sigma\geq \frac{|\kappa|}{2}$, described above, are depicted in Fig. 4.

\begin{figure}[h!]
\centering
\begin{tabular}{cc}
\includegraphics*[width=160pt, height=130pt]{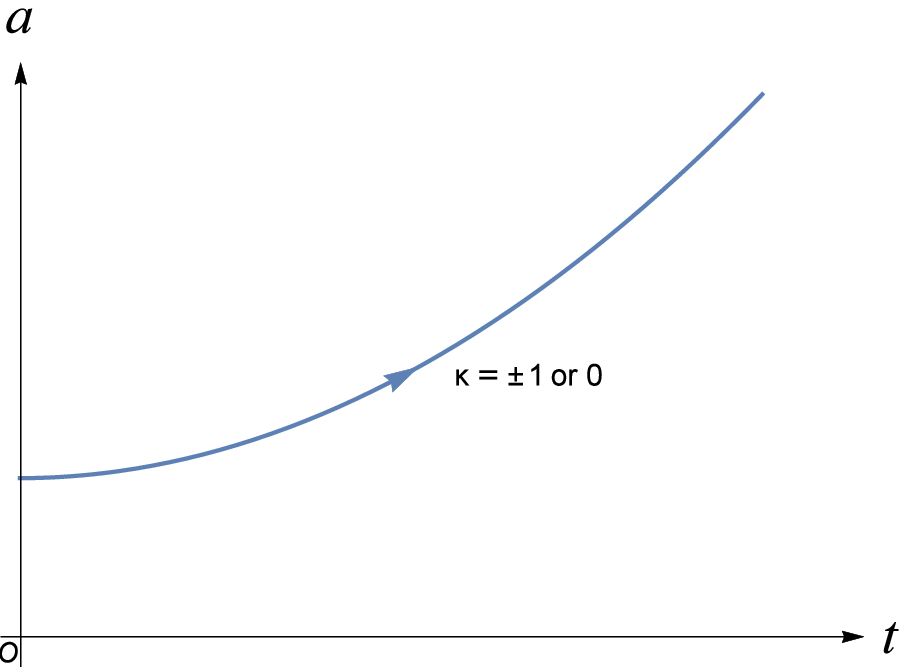}&
\hspace{2cm}
\includegraphics*[width=190pt, height=130pt]{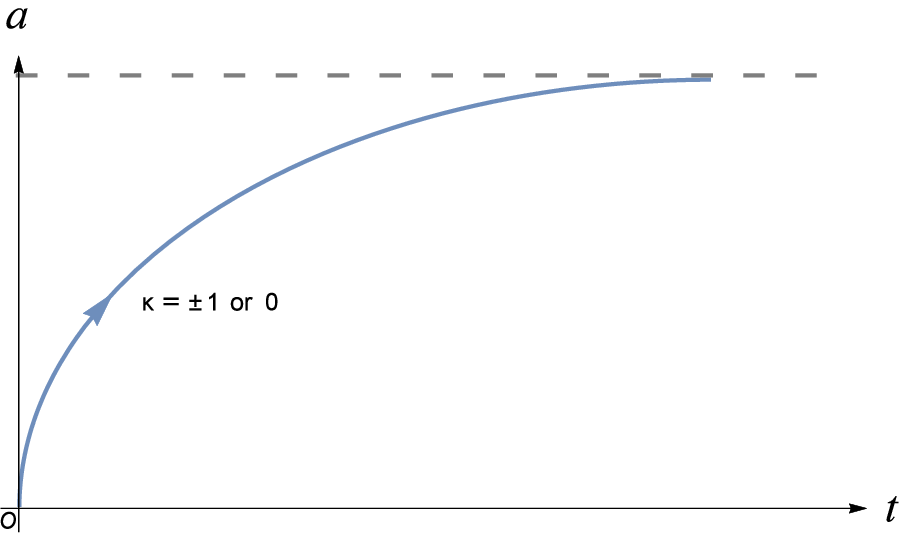}\\
\color{brown}{upper branch}   & \hspace{2cm} \color{brown}{lower branch}
\end{tabular}
\caption{Scale factor evolution for $\omega\sigma\geq \frac{|\kappa|}{2}$.}
\end{figure}

As mentioned, the above two branches (first and second solution) cannot be matched up to a single
extended solution. The reason is that at the end points $a=a_{1}$ and $a=a_{2}$ of the two branches,
the second derivative $\ddot{a}$ is $\frac{\sigma}{a_{1}}$ and $-\frac{\sigma}{a_{2}}$
respectively. This discontinuity from a negative to a positive value does not allow a smooth
matching between the two branches in order to create a smooth cosmic history. The same can be seen
even more explicitly looking at equation (\ref{ikijm}), which is written in terms of the Hubble
parameter as $\frac{dH}{da}=\pm\frac{1}{a^{2}}\sqrt{\omega^{2}a^{2}H^{2}\!-\!\kappa\!+\!
\frac{\sigma^2}{a^{2}H^{2}}}-\frac{H}{a}$. This equation provides that the values of $\frac{dH}{da}$
on the left and right of the candidate matching point are $-\infty$ and $+\infty$
respectively, excluding definitely a smooth matching between the two branches.

$\bullet\,\,\,\,$ $\omega\sigma<\frac{|\kappa|}{2}\,.$
\newline
Here the analysis is different for the two possible cases $\kappa=-1$ and $\kappa=1$ of spatial
topology.

Starting with $\dashuline{\,\textit{$\kappa=-1$}\,}$, where the situation is simpler since there is
only one solution for each $\pm$ branch, it arises that $\epsilon=1$. These solutions are
reminiscent to the solutions above with $\omega\sigma\geq \frac{|\kappa|}{2}\,.$ So, for the upper
branch it arises that
$a\geq a_{1}=a_{o}\big(1\!-\!\frac{\kappa}{2\omega\sigma}\big)^{\!\frac{1}{2\omega}}$, while for the
lower branch it is
$a\leq a_{2}=a_{o}\big(1\!-\!\frac{\kappa}{2\omega\sigma}\big)^{\!-\frac{1}{2\omega}}$.
The {\textit{upper branch (first solution)}} starts expanding from $a=a_{1}$ with $H=0$, $\dot{a}=0$,
$\frac{\ddot{a}}{a}=\frac{\sigma}{a_{1}^{2}}$, up to infinite volume, being always
accelerating; this acceleration could be examined if it helps to the early inflationary era of
the universe. Due to the translation of time origin, the time at the starting point of the
expansion can be chosen to be $t=0$. It can be seen that the Ricci scalar $R$ and the energy density
$\rho$ are finite at the starting point $a=a_{1}$, therefore this solution is non-singular.
The {\textit{lower branch (second solution)}} starts
expanding from $a=0$, where $H$, $\dot{a}$ diverge to $+\infty$, while $\frac{\ddot{a}}{a}$ diverges
to $-\infty$. The initial time can be set to be $t=0$, and close to $t=0$ it is
$H=\frac{\sqrt{\sigma}}{\sqrt{2\omega}\,a_{o}}(\frac{a_{o}}{a})^{\omega+1}$. The expansion continues
until the end point $a=a_{2}$, where $H=0$, $\dot{a}=0$,
$\frac{\ddot{a}}{a}=-\frac{\sigma}{a_{2}^{2}}$, and the evolution is always decelerating.
Also here, the upper and lower branches cannot match smoothly to a unified solution due to the
discontinuity of $\ddot{a}$ or also of $\frac{dH}{da}$. The basic qualitative features of the
possible cosmic evolutions of the two branches with $\omega\sigma<\frac{|\kappa|}{2}$ and
$\kappa=-1$, described above, are depicted in Fig. 5.

\begin{figure}[h!]
\centering
\begin{tabular}{cc}
\includegraphics*[width=160pt, height=130pt]{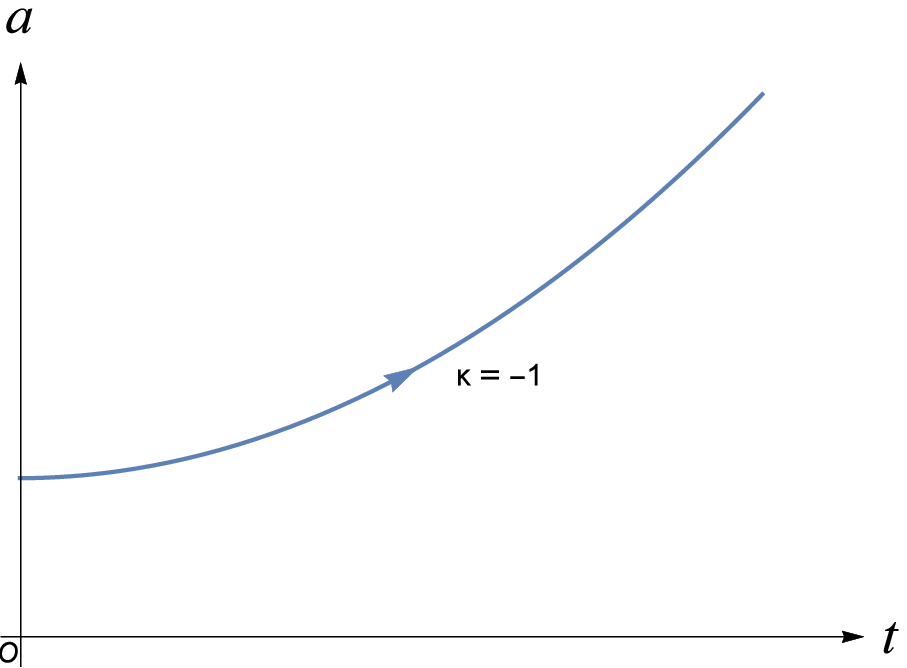}&
\hspace{2cm}
\includegraphics*[width=190pt, height=130pt]{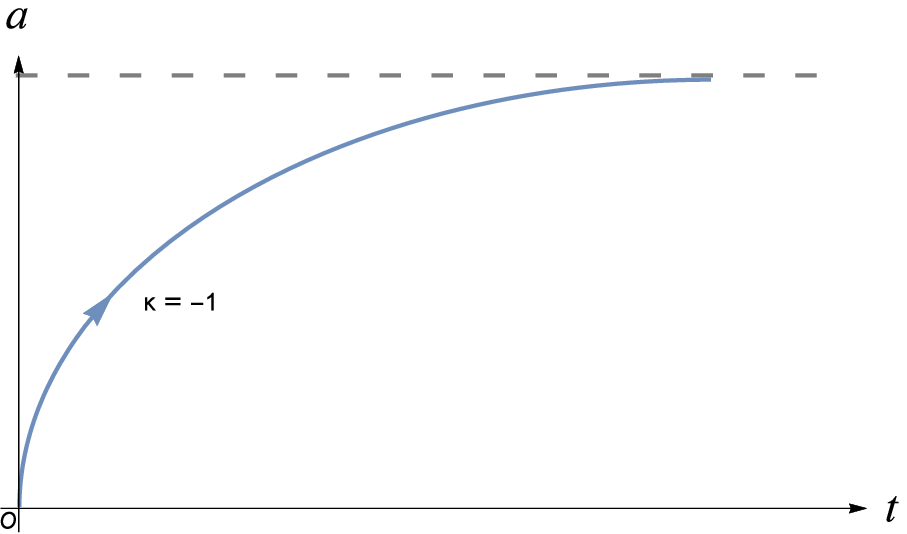}\\
\color{brown}{upper branch}   & \hspace{2cm} \color{brown}{lower branch}
\end{tabular}
\caption{Scale factor evolution for $\omega\sigma< \frac{|\kappa|}{2}$, $\kappa=-1$.}
\end{figure}

We now pass to the analysis of the universe evolution with $\dashuline{\,\,\textit{$\kappa=1$}\,\,}$.
The two $\pm$ branches appeared throughout are due to the presence of $\dot{u}$ squared in equation
(\ref{ikijm}). Although the solutions described above make sense as independent branches and
no matching between different branches can be performed, on the contrary, the branches-solutions
to be analyzed below match smoothly and form two maximally extended physically meaningful solutions.
\newline
We start with the upper $\pm$ branch, where there are now two different solutions. The first one has
$a\geq a_{3}=a_{o}\big(\frac{\kappa^{2}}{4\omega^{2}\sigma^{2}}\!-\!1\big)^{\!\frac{1}{4\omega}}$
and $\epsilon=1$. This solution describes an expanding, always accelerating, universe which starts
from $a=a_{3}$ with $H=\frac{\sqrt{\sigma}}{\sqrt{\omega}\,a_{3}}
\Big(\!\sqrt{\!\frac{\kappa^{2}}{4\omega^{2}\sigma^{2}}\!-\!1}\!+\!
\frac{\kappa}{2\omega\sigma}\!\Big)^{\!1/2}$, $\ddot{a}=0$, and continues up to infinite volume;
this acceleration could be examined if it helps to the early inflationary era of the universe.
The second expanding solution occurs for $a_{1}'\leq a\leq a_{3}$, where
$a_{1}'=a_{o}\big(\frac{\kappa}{2\omega\sigma}\!-\!1\big)^{\!\frac{1}{2\omega}}$, and it has
$\epsilon=-1$. It starts from $a=a_{1}'$ with $H=0$, $\frac{\ddot{a}}{a}=\frac{\sigma}{a_{1}'^{2}}$,
it is always accelerating and ends at $a=a_{3}$ with $H=\frac{\sqrt{\sigma}}{\sqrt{\omega}\,a_{3}}
\Big(\!\frac{\kappa}{2\omega\sigma}\!-\!
\sqrt{\!\frac{\kappa^{2}}{4\omega^{2}\sigma^{2}}\!-\!1}\Big)^{\!1/2}$, $\ddot{a}=0$;
this acceleration could also be examined if it helps to the early inflationary era of the universe.
It can be seen that the Ricci scalar $R$ at the starting point $a=a_{1}'$ is finite; moreover, from
equation (\ref{djbng}) the energy density $\rho$ at $a_{1}'$ is also finite, therefore this solution
is clearly non-singular. Although this branch will be matched below with another branch to form
the full solution, however the non-singular point $a_{1}'$ will remain intact as the one
non-singular endpoint of the total evolution.
\newline
We continue with the analysis of the lower $\pm$ branch, where still there are two
different solutions. The first of these solutions has $a\leq a_{4}=a_{o}
\big(\frac{\kappa^{2}}{4\omega^{2}\sigma^{2}}\!-\!1\big)^{\!-\frac{1}{4\omega}}$ and
$\epsilon=1$. This solution starts expanding from $a=0$, where $H$, $\dot{a}$ diverge
to $+\infty$, while $\frac{\ddot{a}}{a}$ diverges to $-\infty$. The initial time can be set to be
$t=0$, and close to $t=0$ it is
$H=\frac{\sqrt{\sigma}}{\sqrt{2\omega}\,a_{o}}(\frac{a_{o}}{a})^{\omega+1}$. The expansion continues
until the end point $a=a_{4}$, where $H=\frac{\sqrt{\sigma}}{\sqrt{\omega}\,a_{4}}
\Big(\!\sqrt{\!\frac{\kappa^{2}}{4\omega^{2}\sigma^{2}}\!-\!1}\!+\!
\frac{\kappa}{2\omega\sigma}\!\Big)^{\!1/2}$, $\ddot{a}=0$, and the evolution is always decelerating.
The second expanding solution occurs for $a_{4}\leq a\leq a_{2}'$, where
$a_{2}'=a_{o}\big(\frac{\kappa}{2\omega\sigma}\!-\!1\big)^{\!-\frac{1}{2\omega}}$, and it has
$\epsilon=-1$. It starts from $a=a_{4}$ with
$H=\frac{\sqrt{\sigma}}{\sqrt{\omega}\,a_{4}}\Big(\!\frac{\kappa}{2\omega\sigma}\!-\!
\sqrt{\!\frac{\kappa^{2}}{4\omega^{2}\sigma^{2}}\!-\!1}\Big)^{\!1/2}$, $\ddot{a}=0$, it is
always decelerating and it ends at $a=a_{2}'$ with $H=0$,
$\frac{\ddot{a}}{a}=-\frac{\sigma}{a_{2}'^{2}}$.
\newline
After having described the various branches for $\kappa=1$, we now come to the appropriate matchings
in order to define the possible total solutions. Each branch has an integration constant $a_{o}$.
However, the differential equation (\ref{ikijm}) is only of first order, so only a single
integration constant $a_{o}$ should describe a total solution of equation (\ref{ikijm})
instead of the two different $a_{o}$'s of the two branches which are matched together.
Therefore, the following two total solutions are the only ones which arise, the first with
$\epsilon=1$ and the second with $\epsilon=-1$.
\newline
\textit{Total solution 1.} The first solution described above (which we call here with respect
to the time arrow as left) of the lower branch matches naturally and smoothly with the
first solution (named as right) of the upper branch. Namely, if $a_{o-}$ and $a_{o+}$
denote the two corresponding integration constants of the left and right solutions
respectively and it holds $a_{o+}=a_{o-}
\big(\frac{\kappa^{2}}{4\omega^{2}\sigma^{2}}\!-\!1\big)^{\!-\frac{1}{2\omega}}$, then the two
solutions match at $a_{4}=a_{3}$. Due to the existence of time translation invariance, the
matching of the corresponding times is trivial. At the matching point, as explained above, the
values of $H$, $\ddot{a}$ (and thus also $\dot{H}$) coincide for the left and right solution.
These results are sufficient to guarantee the full smoothness of $a(t)$ since progressive higher
differentiations of (\ref{hyd}) generate all the higher derivatives of $H$ only at linear order;
therefore, these derivatives can uniquely be solved algebraically at the matching point in terms of
lower derivatives, and as a result, these higher derivatives are also continuous. Since all the
higher derivatives $a^{(n)}$ of $a$ are expressed in terms of higher derivatives of $H$, thus
$a^{(n)}$ are also continuous at the matching point. This means that the full solution, consisting
of the two separate branches, is infinitely smooth. This total solution undertakes a continuous
expansion from $a=0$ up to infinite volume and the solution is initially decelerating and
finally accelerating. The matching of the two branches can also be seen directly from equation
(\ref{ikijm}), which is written in terms of the Hubble parameter as
$\frac{dH}{da}=\pm\frac{1}{a^{2}}\sqrt{\omega^{2}a^{2}H^{2}\!-\!\kappa\!+\!
\frac{\sigma^2}{a^{2}H^{2}}}-\frac{H}{a}$, without reference to time. This equation provides that
the values of $\frac{dH}{da}$ on the left and right of the matching point are both $-\frac{H}{a}$,
and therefore coincide. All higher derivatives also match after successive differentiations of
(\ref{ikijm}).
\newline
\textit{Total solution 2.} The second solution (left one) of the upper branch and the second
solution (right one) of the lower branch also match together. Now, the integration constant of the
right branch should be $a_{o+}=a_{o-}
\big(\frac{\kappa^{2}}{4\omega^{2}\sigma^{2}}\!-\!1\big)^{\!\frac{1}{2\omega}}$ and the matching
occurs at $a_{3}=a_{4}$. Also here, at the matching point, the solution is infinitely smooth, so
all derivatives of $a(t)$ coincide for the two branches. For the total cosmic evolution, the
scale factor starts from $a=a_{1}'$, where the solution is non-singular (in Ricci scalar and
energy density) and accelerating, and later there is a passage into deceleration until the end
point $a=a_{2}'$. This initial transitory accelerating era may be of relevance to the exit from
inflation question in the early universe. The smooth matching of the two branches can also be seen
directly from equation (\ref{ikijm}), without reference to time, providing that the values of
$\frac{dH}{da}$ on the left and right of the matching point are both $-\frac{H}{a}$, and therefore
coincide; all higher derivatives also match after successive differentiations of (\ref{ikijm}).
\newline
The basic qualitative features of the possible cosmic evolutions of the two total solutions with
$\omega\sigma<\frac{|\kappa|}{2}$ and $\kappa=1$, described above, are depicted in Fig. 6.

\begin{figure}[h!]
\centering
\begin{tabular}{cc}
\includegraphics*[width=180pt, height=130pt]{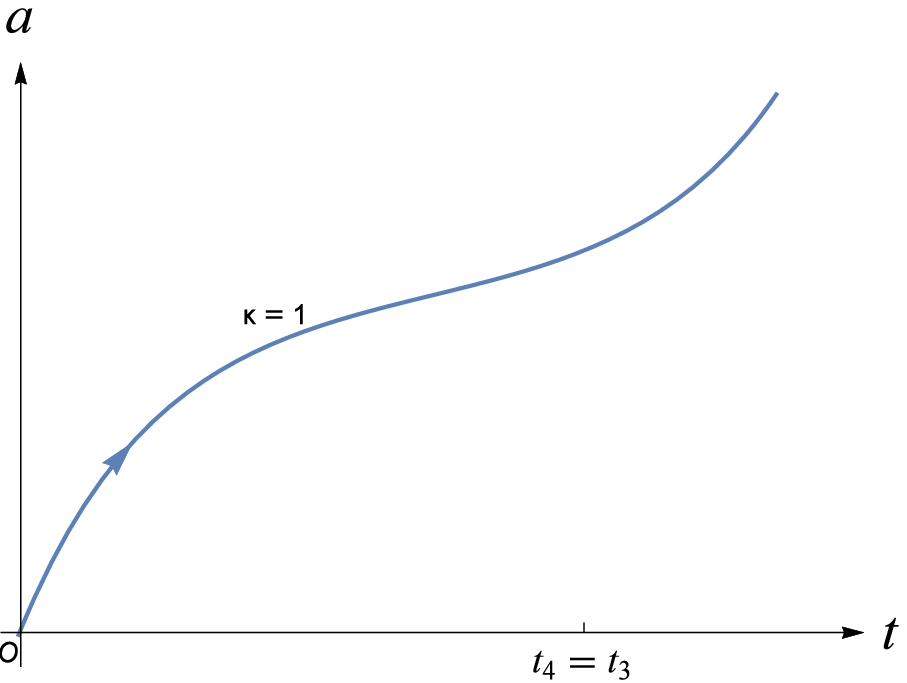}&
\hspace{2cm}
\includegraphics*[width=190pt, height=130pt]{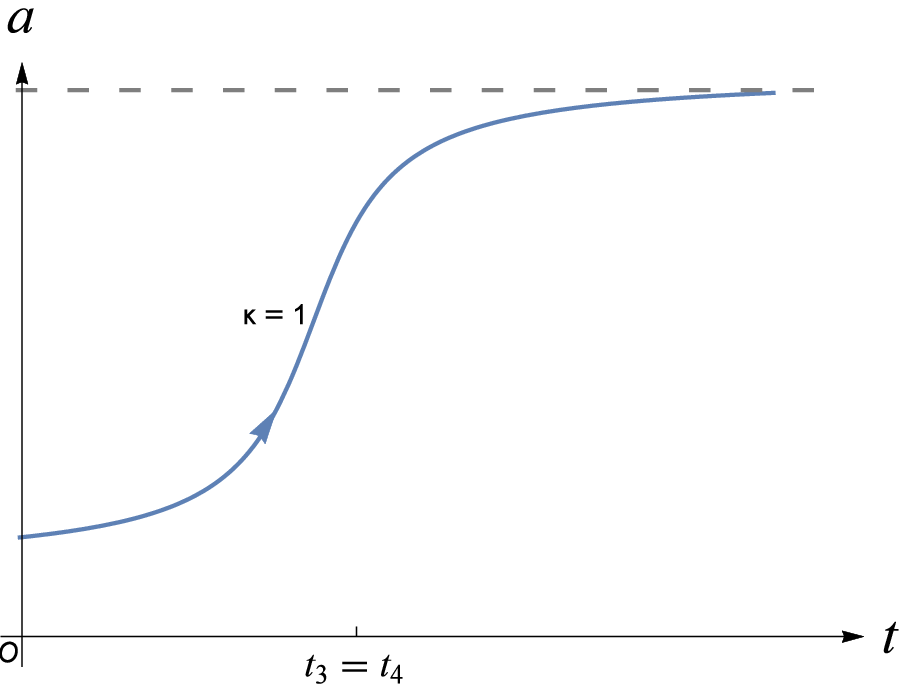}\\
\color{brown}{total solution 1}   & \hspace{2cm} \color{brown}{total solution 2}
\end{tabular}
\caption{Scale factor evolution for $\omega\sigma< \frac{|\kappa|}{2}$, $\kappa=1$.}
\end{figure}

Note also that all solutions obtained in this subsection with $k\propto H$ have an infinite
Ricci scalar $R$ whenever the scale factor $a$ vanishes. In this sense, the expanding solutions
which start at $a=0$ with a big bang or the contracting solutions which result to $a=0$ with a big
crunch can be considered as singular with respect to the divergences of the curvature invariants.

To summarize very briefly the behaviour of all the expanding solutions found above for the scaling
$k\propto H$ : (i) there exist (for all spatial curvatures) non-singular (in all scale factor, Ricci
scalar and energy density) accelerating solutions growing from a finite scale factor to infinity,
(ii) there are (for all spatial curvatures) solutions evolving from a big bang to a finite scale
factor, (iii) there are (for positive spatial curvature) solutions between zero and infinite scale
factor, and finally, (iv) there exist (for positive spatial curvature) non-singular (in all scale
factor, Ricci scalar and energy density) solutions, evolving between two different finite scale
factors and possessing an initial transient acceleration. The time reversal of these expanding
solutions provides the corresponding collapsing solutions.

\section{Discussion and Conclusions}
\label{ekjdk}

Last decades physicists explored the possibility that the coupling constants of certain field
theories might become functions of spacetime coordinates. In gravity there is a long time research
concerning spacetime varying Newton and cosmological ``constants'' $G(x)$ and $\Lambda(x)$.
As yet, no well established formalism has been developed towards a generic running of these
coupling constants in a modification of standard Einstein's relativity, working with a Lagrangian
or Hamiltonian approach.

The present work follows an alternative approach. Working at the level of equations of motion,
mathematically and physically consistent, gravitational equations in the presence of matter
with spacetime dependent $G(x)$ and $\Lambda(x)$ have been derived by making an exhaustive analysis
of the Bianchi identities. These novel equations are unique and the most general ones under the
assumption of containing up to second derivatives in the metric and $G,\Lambda$. In the modified
Einstein equations, only kinetic terms of the cosmological constant arise, while all the possible
kinetic terms of Newton constant vanish. However, in the modified conservation equation of matter,
not only terms with derivatives of $\Lambda$, but also derivatives of $G$ are present. The case
where the Newton constant is of the form $G(\Box)$, and acts as a derivative operator on the
energy-momentum tensor, is also included in our formulation.

We choose the Asymptotic Safety scenario of quantum gravity as a concrete example to
perform an explicit application of this new formalism. In the high energies regime, the
non-Gaussian fixed point is well established in the context of Asymptotic Safety and provides the
form of $\Lambda,G$ in terms of the energy scale $k$. The early universe is the corresponding
appropriate era during the cosmological evolution where we carried out the analysis. We have chosen
various typical scaling laws relating $k$ to spacetime, and so, specific functions $\Lambda(x),G(x)$
appear in terms of cosmological parameters.

The general cosmological solutions for all spatial curvatures have been derived, assuming a perfect
fluid with a radiation equation of state. It is interesting that, although the new effective field
equations contain extra terms due to the running of the constants, the energy density of radiation
fluid obeys the standard scaling law $a^{-4}$ in terms of the scale factor $a$, what is interesting
since this law is important for the successful description of the early thermal history of the
universe. There is a variety of cosmic evolutions obtained, which exhibit phenomenologically
interesting behaviours, depending on the parameter values, the spatial topology and the selected branch.
To briefly emphasize only on a few such behaviours: (i) there are bouncing solutions (for positive
spatial curvature) from infinite to infinite scale factor, (ii) there are recollapsing solutions
(for positive spatial curvature) which start from a big bang with an initial transient accelerating
era and end up to a big crunch, (iii) there exist (for all spatial curvatures) non-singular (in all
scale factor, Ricci scalar and energy density) expanding and accelerating solutions growing from a
finite scale factor to infinity, and finally, (iv) there are (for positive spatial curvature)
non-singular (in all scale factor, Ricci scalar and energy density) expanding solutions, evolving
between two different finite scale factors and possessing an initial transient acceleration.
There are solutions (such as the above mentioned or others) which are accelerating or possess only
a temporary accelerating period; it would be interesting to examine if these solutions are capable
to describe appropriately the inflationary features of the early universe, or even if they can
provide a natural exit from inflation already inside the high energy radiation regime.

\begin{acknowledgments}

The authors acknowledge the support of Orau small grant SOE2019010 ``Quantum Gravity at
astrophysical distances''.

\end{acknowledgments}

\end{document}